\documentclass[ reprint,
amsmath,amssymb,
aps,
]{revtex4-2}

\usepackage{epsf,epsfig}
\usepackage{amssymb,amsmath,amsfonts}
\usepackage{color}
\usepackage{braket}
\usepackage{subcaption}
\usepackage[T1]{fontenc} 

\usepackage[dvipsnames, svgnames,  x11names ]{xcolor}
\usepackage[colorlinks]{hyperref}
\hypersetup{
   citecolor = {blue},
   linkcolor = {RubineRed},
}

\usepackage{physics}
\usepackage{tikz}
\usetikzlibrary{arrows,snakes,backgrounds}

\usepackage[english]{babel}

\usepackage{graphicx}

\definecolor{NordCyan}{HTML}{8FBCBB}          
\definecolor{NordBrightCyan}{HTML}{88C0D0}    
\definecolor{NordBlue}{HTML}{81A1C1}          
\definecolor{NordBrightBlue}{HTML}{5E81AC}    
\definecolor{NordRed}{HTML}{BF616A}           
\definecolor{NordOrange}{HTML}{D08770}        
\definecolor{NordYellow}{HTML}{EBCB8B}        
\definecolor{NordGreen}{HTML}{A3BE8C}         
\definecolor{NordMagenta}{HTML}{B48EAD}       

\begin{document}

\title{Krylov complexity for 1-matrix quantum mechanics}


\author{Niloofar Vardian}
\email{nvardian@sissa.it}
\affiliation{SISSA, International School for Advanced Studies, via Bonomea 265, 34136 Trieste, Italy\\
}

\begin{abstract}
\noindent
This paper investigates the notion of Krylov complexity, a measure of operator growth, within the framework of 1-matrix quantum mechanics (1-MQM). Krylov complexity quantifies how an operator evolves over time by expanding it in a series of nested commutators with the Hamiltonian. We analyze the Lanczos coefficients derived from the correlation function, revealing their linear growth even in this integrable system. This growth suggests a link to chaotic behavior, typically unexpected in integrable systems. Our findings in both ground and thermal states of 1-MQM provide new insights into the nature of complexity in quantum mechanical models and lay the groundwork for further studies in more complex holographic theories.

\end{abstract}
\maketitle
\section{Introduction}

The notion of “complexity” is playing an increasingly important role in several physical contexts \cite{nielsen2010quantum}, from computational condensed matter to holographic spacetime \cite{susskind2020three}.
This quantity should reflect how "complicated" a physical system is. In quantum mechanics, this complexity could relate to states compared to a basic reference state, to operators, or a mix of both \cite{rabinovici2022krylov}.
In other words, this idea of growth is connected to how we understand complexity in quantum systems. Essentially, complexity measures how difficult it is to create a state from a basic starting point. There are two main ways to look at complexity:
\begin{itemize}
\item How complex a state is.
\item How an operator spreads out over time.
\end{itemize}
Recently, there have been many discoveries in these areas \cite{balasubramanian2022quantum,  caputa2022quantum, bhattacharjee2023operator, caputa2023spread, afrasiar2023time, alishahiha2023quantum, Vasli:2023syq, Alishahiha:2022anw, bhattacharjee2022probing}. 
For more details, you can look at studies on quantum systems 
\cite{jefferson2017circuit, chapman2019complexity, caceres2020complexity, chagnet2022complexity}, 
quantum field theories 
\cite{stanford2014complexity,susskind2020three, susskind2018things, chattopadhyay2023spread}, 
black hole physics  
\cite{chapman2022quantum, muck2022krylov},
 as recent reviews of these topics \cite{nandy2024quantum}.
The main idea is to describe how entangled a state is using simple parts, like gates in a quantum circuit or tensor network. It is used for the state complexity, and then complexity is measured by the size of the smallest circuit that can represent the state using these parts. For more details, see \cite{nielsen2005geometric}.

Recent work has shifted focus from states to operator growth in many-body systems 
\cite{von2018operator,
chan2018solution}.
The authors of reference \cite{parker2019universal} introduced a new way to understand complexity by looking at how operators evolve over time. Instead of using a fixed set of operators, they use a basis that changes with time. Starting with an initial operator 
$O_0$
they consider how it evolves in the Heisenberg picture, given by
$ O(t) = e^{iHt} O_0 e^{-iHt}$ .
 This evolution can be expanded into a series using nested commutators with the Hamiltonian. These nested commutators act as building blocks for describing how the operator evolves. The concept of K-complexity from \cite{parker2019universal} measures the "effective dimension" or the size of the space that the evolving operator explores.
Krylov space is defined as the linear span of nested commutators$ [H . . . , [H, O]]$, where $H$ is the system’s Hamiltonian and $O$ is an operator in question.
More precisely, the operator may be expanded in terms of nested operators as follows
\begin{equation}
O(t) = e^{iHt} O_0 e^{-iHt}= \sum_{0}^\infty \frac{(it)^n}{n!} \mathcal{L}_n O
\end{equation}
where $ \mathcal{L}_n O = \{O, [H,O], [H,[H,O]],...\}$. 
Given a proper inner product in the space of operators, these nested operators are not orthogonal or normalized. However, we can construct an orthogonal and normalized basis called the Krylov basis. This is done using the Gram-Schmidt process.

Starting with an autocorrelation function  $C(t)$
of a simple local operator $O$,  you can use the recursion method to find Lanczos coefficients $b_n$.
These coefficients describe how the operator grows in the Krylov subspace. The original work \cite{parker2019universal} suggested the universal operator growth hypothesis, which links the long-term behavior of $b_n$ to the dynamics of the system.
There is non-trivial evidence supporting the connection between the behavior of $b_n$
and integrability/chaos, yet it does not seem to be universal \cite{bhattacharjee2022krylov}.
In \cite{Craps:2023ivc}, authors explore the possible connection of the Krylov complexity with the circuit complexity.

We investigated aspects of Krylov complexity in a system of 1-matrix quantum mechanics (1-MQM). 
 The notion of Krylov complexity has the advantage that it is well-defined for the class of quantum mechanical theories which appear in the context of the AdS/CFT correspondence, in particular large N gauge theories in various dimensions. We explored aspects of Krylov complexity in the model of quantum mechanics of a single Hermitian matrix.
Using the mapping at large N to a gas of non-interacting fermions moving in an external potential We are  able to compute the 2-point function of single-trace operators in this model
and from that extract the asymptotic form of the so-called Lanczos coefficients. While the system is integrable these coefficients display linear growth which in the literature has been
conjectured to be related to chaotic features. The work is an important first step towards the goal of computing Krylov complexity in more realistic holographic theories,
involving more than one matrix, for example, the BFSS matrix model.

We study the notion of Krylov complexity in both the ground state and thermal state. 
In the ground state, we see the linear growth of the Lancsoz coefficients albeit the theory is integrable.  In the thermal state, we find that the Lancsoz coefficients contain
the even and odd linear branches. Moreover, we will find the radius of convergence when
we are using the correlator to find the Krylov complexity. Till that point, we see only growth of the Krylov
complexity. It is almost the same point as the first peak of the correlation
function.

The structure of the paper is as follows: 
In Chapter 2, we review the concept of the Kyrilov complexity. After that in Chapter 3, we study the behavior of the Krylov complexity and Lancsoz coefficients for the $J$ number of decoupled harmonic oscillators. In Chapter 4, we review the basics of the 1-MQM and find the correlators of the theory both in the ground state and thermal one.  In Chapter 5, we discuss the notion of complexity in 1-MQM. And in the end, in Chapter 6, we discuss the radius of convergence of the Krylov complexity.

\section{Krylov Complexity}
We start with the definition of the notion of the Krylov complexity. It is defined as the recursion method in \cite{viswanath1994recursion} and recently has been used in \cite{parker2019universal}.

\subsection{Krylov state complexity}

Consider a quantum system with a time-independent Hamiltonian $H$. A state $ \ket{\psi (t)}$ is time evolved under the Schrodinger equation $ i \partial_t \ket{\psi (t)} = H \ket{\psi(t)}$. Its solution $ \ket{\psi(t)} =e^{-i H t} \ket{\psi(0)}$ 
has a formal power series expansion 
\begin{equation}
    \ket{\psi(t)} = \sum _{n=0}^\infty \frac{(it)^n}{n!} \ket{\psi_n}
\end{equation}
while $ \ket{\psi_n} = H^n \ket{\psi(0)}$. The time-evolved state  is a linear combination of 
\begin{equation}\label{basis}
    \ket{\psi(0)}, \qquad \ket{ \psi_1} = H \ket{\psi(0)}, \qquad \ket{\psi_2} = H^2\ket{\psi(0)},~ ...~.
\end{equation}
The subspace $ \mathcal{H}_{\psi}$ which is spanned by \eqref{basis} is called \emph{Krylov subspace}. Notice that in general, this basis is not orthogonal.
The Gram-Schmidt procedure applied to $ \ket{\psi_n}$ generate an orthogonal basis 
$ \mathcal{K}= \{ \ket{K_n}: n=0,1,2,..., K_\psi\}$ when we define $ K_\psi = \dim \mathcal{H}_\psi$ for one subspace of the full Hilbert space explored by the evolution of $ \ket{\psi(0)} = \ket{K_0}$. In general, this code subspace can be infinite dimension.

Using the ordinary inner product, one can orthogonalize the basis \eqref{basis} through the \emph{Lanczos algorithm}:

\begin{enumerate}
    \item  $b_0 \equiv 0, \qquad\ket{K_{-1}}=0$ 
    \item $ \ket{K_0} \equiv \ket{\psi(0)}, \qquad a_0 = \bra{K_0} H \ket{K_0}$
    \item  For $n\geq 1$, $\ket{A_n}= (H-a_{n-1}) \ket{K_{n-1}} - b_{n-1} \ket{K_{n-2}}$
    \item  Set $ b_n = \sqrt{\langle A_n| A_n \rangle}$
    \item If $b_n =0 $ stop, otherwise set $ \ket{K_n}= \frac{1}{b_n} \ket{A_n}$, $ a_n = \bra{K_n} H \ket{K_n}$ and go to step 3.
    
\end{enumerate}
In the case that $ K_\psi$ is finite, the Lanczos algorithm will end at some point that $ b_{K_\psi} =0$. The result of the Lanczos algorithm is two sets of Lanczos coefficients $\{a_n\}$ and $ \{b_n\}$. 

We can expand the time-evolved state in terms of the Krylov basis as 
\begin{equation}
    \ket{\psi(t)} = \sum _{n=0}^{K_\psi -1} \phi_n(t) \ket{K_n}
\end{equation}
by substituting it into the Schrodinger equation, one gets
\begin{equation}
    \dot{\phi}_n(t) = a_n \phi _n(t) + b_{n+1} \phi_{n+1}(t) + b_n \phi _{n-1}(t)
\end{equation}
and the initial condition is $ \phi_n(0) = \delta _{n,0}$.

The Krylov state complexity of the state $ \ket{\psi (t)}$ is defined as 
\begin{equation}
    C_\psi (t) \equiv \sum _{n=0} ^{K_\psi -1} n | \phi_n(t)|^2.
\end{equation}

\subsection{Krylov operator complexity}

Similar to the Krylov state complexity, we can define Krylov complexity for quantum operators. 
Motivated by the time evolution of the operators , one can create the Krylov basis for a given operator in terms of the nested commutators with the Hamiltonian as they determine the time Taylor expansion of the Heisenberg operator. 

Consider a time-independent Hamiltonian of a quantum system $H$ and a given Hermitian operator $O$. The operator undergoing a Heisenberg evolution
\begin{equation}
    O(t) = e^{itH} O(0) e^{-itH}.
\end{equation}
Just as states evolved under the Hamiltonian operator, operators evolved under the Liouvillian operator 
$ \mathcal{L}\equiv [H,.]$
\begin{equation}
    \begin{split}
        O(t) =& e^{itH} O(0) e^{-itH} = O(0) + it [H, O(0)] + ...
        \\
        =& \sum _{n=0}^\infty \frac{(it)^n}{n!} \mathcal{L}^n O(0) \equiv e^{i\mathcal{L}t} O(0)
    \end{split}
\end{equation}
This is a linear combination of the sequence of operators 
\begin{equation}\label{basis2}
    O, \qquad
    \mathcal{L}  O = [H,O],\qquad
     \mathcal{L}^2  O = [H,[H,O]],\qquad
     ...
\end{equation}
where $O$ stands for $ O(0)$ \cite{kundu2023state}.
The linear span of operators forms an invariant subspace $ \mathcal{H}_O$. 
A convenient way to study the growth of a simple operator is to realize them as states, $ O\equiv \ket{O}$, and to introduce a notion of an inner product. It can be any non-degenerate inner product in the operator algebra such as the trace inner product for finite-dimensional Hilbert space (also known as infinite temperature inner product or Frobenius norm)
\begin{equation}
    \langle O| O' \rangle = \frac{\Tr [O^\dagger O']}{\Tr[I]}
\end{equation}
and we write $ ||O|| =\langle O|O\rangle ^{1/2} $ for the norm \cite{bhattacharya2023krylov}.
Thereby any operator within this subspace can be thought of as a vector in the linear vector space. Such a vector space endowed with a valid inner product is called the \emph{Krylov subspace}.

The set of operators \eqref{basis2}  are not orthogonal. The idea is to apply the Gram-Schmidt to orthogonalize it. As in the case of the state complexity, it is called Lanczos algorithm. It is as follows
\begin{enumerate}
    \item $b_0 \equiv 0, ~~ O_{-1} \equiv 0$
    \item  $ O_0 = O /||O||$
    \item For $ n \geq 1$ : $ A_n = \mathcal{L} O_{n-1} - b_{n-1} O_{n-2}$
    \item Set $ b_n = ||A_n||$
    \item  If $ b_n =0 $ stop; otherwise set $ O_n= A_n /b_n$ and go to step 3.
\end{enumerate}
The output of the algorithm is a sequence of positive numbers, $\{b_n\}$, called the Lanczos coefficients and 
an orthogonal set of operators $\{ O_n\}_{n=0}^{K_O-1}$ called the Krylov basis.

The time-evolved operator can now be expanded on the Krylov basis
\begin{equation}
    O(t) = e^{iHt} O_0 e^{-iHt} = \sum _{n=0}^{K_O -1} i^n \phi_n (t) O_n
\end{equation}
where $ \phi_n(t)$ can be thought of as the wavefunction over the Krylov basis.
From the orthogonality, we obtain
\begin{equation}
    \phi_n(t) =  i^{-n} \langle O_n | O(t) \rangle.
\end{equation}
The time evolution of the operator follows
\begin{equation}
    \begin{split}
       \frac{d O(t)}{dt}& = \sum _n i^n \frac{d\phi_n(t)}{dt} O_n
       \\
       &= i[H, O(t)] = i \mathcal{L} O(t) = \sum_n i^{n+1} \phi_n(t) \mathcal{L}O_n
    \end{split}
\end{equation}
thus via the Heisenberg equation $ \phi_n(t)$ satisfies the equation
\begin{equation}
    \partial_t \phi_n(t) = b_n \phi_{n-1}(t) - b_{n+1} \phi_{n+1}(t)
\end{equation}
with boundary condition $ \phi_{-1} (t) =0$ and $ \phi_n(t=0) = \delta_{0,n}$.
From unitarity, since the initial operator is normalized at the first step of the Lanczos algorithm, the wavefunction $ \phi_n(t)$ is normalized at all times 
$ \sum_{n=0} ^{K_O -1} | \phi_n(t)|^2 =1$.

\emph{Krylov complexity} or K-complexity is defined as the time-dependent average position over the Krylov chain 
\begin{equation}\label{22222}
    C_K (t) = \bra{O(t)} n \ket{O(t)} = \sum_n n |\phi_n(t)|^2
\end{equation}
which can be viewed as the expectation value of the Krylov operator
\begin{equation}
    K_O = \sum_n n \ket{O_n} \bra{O_n}.
\end{equation}
Intuitively, $C_K(t)$ describes the mean width of a wavepacket in the Krylov space and hence quantitatively measures how the size of the operator increases as time goes by \cite{lv2023building}.

\subsection{Krylov operator complexity over pure and mixed states}

Given an normalized operator $O$, by acting the operator on a pure state $ \ket{\psi}$, one can construct a state
\begin{equation}
    \ket{O} := O \ket{\psi} \qquad \mathcal{L}^n \ket{O} := [H,[H, ...[H,O]]]\ket{\psi}.
\end{equation}
The choice of pure state $\ket{\psi}$ depends on the two-point function of the operator that we have in hand. For example, when we have the zero-temperature two-point function of $O$, one can take the state $\ket{\psi}$ to be the ground state of the theory.

A time-dependent state $ \ket{O(t)} := O(t) \ket{\psi} $ for $ O(t) = e^{\mathcal{L} t } O $
can be expanded by $ \mathcal{L}^n \ket{O}$. Although they do not create on an orthonormal basis. We need to apply the Gram-Schmidt procedure to make them orthogonal. By using  it, one can obtain the Krylov basis $\ket{O_n}$ such that $ \langle O_m \ket{O_n} = \delta_{m,n}$ as follows
\begin{equation}\label{Arnoldi}
    \ket{O_0} = \ket{O} \qquad  \mathcal{L} \ket{O_n} = \sum _{i=0}^{n+1} h_{i,n} \ket{O_i}.
\end{equation}
This construction of the basis is called \emph{Arnoldi iteration} for general matrices. If $ \bra{O_m} \mathcal{L} \ket{O_n}$ is a Hermitian matrix, then \eqref{Arnoldi} is simplified as
\begin{equation}
    \mathcal{L} \ket{O_n} = a_n \ket{O_n} + b_n \ket{O_{n-1}}+ b_{n+1}\ket{O_{n+1}}
\end{equation}
\begin{equation}
    \bra{O_n} \mathcal{L} \ket{O_m} = \begin{pmatrix}
        a_0 & b_1 & 0&0& \dots
        \\
        b_1& a_2& b_2 & 0& \dots
        \\
        0& b_2& a_2 & b_3 &\dots
        \\
        0&0& b_3 & a_3 &\dots
        \\
        \vdots & \vdots & \vdots & \vdots & \ddots
    \end{pmatrix}
\end{equation}
while $ \ket{O_{-1}} =0$. As before, this construction is called the Lanczos algorithm. 

If $ \ket{\psi}$ is an eigenstate of $H$, let us say $ H \ket{ \psi} = \lambda \ket{\psi}$, we have
\begin{equation}
    \bra{O_m} \mathcal{L} \ket{O_n} = \bra{\psi} O_m^\dagger (H- \lambda ) O_n \ket{\psi}
\end{equation}
which is Hermitian. Assuming  that $O$ and $H$ are Hermitian and we have an appropriate inner product by trace and Hermitian conjugation, we find that 
\begin{equation}
    a_n =0.
\end{equation}
$a_n$ is the Hamiltonian eigenvalue in the absence of $b_n$, which would be not directly related to the spreads of operators. 
The non-trivial existence of $ a_n$
coefficients in physical systems first arises in \cite{bhattacharya2022operator} 
 which is later analytically computed in \cite{bhattacharjee2024operator}.
On the other hand, $b_n$, especially at large $n$, represents how much the operator spreads into an orthogonal direction in the Hilbert space at a later time.

By introducing an inner product between operators at finite temperature one can generalize the above procedure
\begin{equation}
    \langle A \ket{B}_\beta := \frac{1}{Z} \Tr \big(e^{-\beta H } A^\dagger B\big), \qquad\qquad Z = \Tr\big(e^{-\beta H} \big)
\end{equation}
where $ \beta $
 is the inverse temperature. We define
 \begin{equation}
     \bra{A} \mathcal{L}^n \ket{B}_\beta := \langle A \ket{\mathcal{L}^n B}_\beta  = \langle \mathcal{L}^n A \ket{B}_\beta.
 \end{equation}
Once the inner product is defined, one can construct the Krylov basis as $ \langle O_m \ket{O_n} = \delta _{m,n}$. On top of it 
\begin{equation}
    L_{mn} = \langle O_m | \mathcal{L} | O_n \rangle = \frac{1}{Z} \Tr [ e^{-\beta H} (O_m^\dagger H O_n - O_m^\dagger O_n H)]
\end{equation}
which is Hermitian. Hence, one can use the Lanczos algorithm instead of the Arnoldi iteration.
For mixed states, it is more convenient that define the Lanczos coefficients in terms of operators as
\begin{equation}
    \begin{split}
        &O_{-1} =0 , \qquad O_0 = O 
        \\
        & \mathcal{L} O_n = a_n O_n + b_n O_{n-1} + b_{n+1} O_{n+1}.
    \end{split}
\end{equation}
One can also obtain 
\begin{equation}
    \bra{O_m} \mathcal{L}^n \ket{O_n} = (L ^n)_{mn}.
\end{equation}
In the zero temperature limit, this reduces to the Lanczos algorithm for the pure state case  
when $ \ket{\psi}$ is the ground state of the theory and in the infinite temperature limit, it reaches the discussion of the Krylov operator complexity \cite{iizuka2023krylov}.

\subsection{Recursion Method and Moment Expansion}
This part is mostly based on \cite{viswanath1994recursion}.



The dynamical behavior of a quantum system is determined by its Hamiltonian $H$ and the operator $ A$ representing the observable we're interested in tracking over time. Our objective is to compute the dynamical correlation function $ \langle A(t) A(0) \rangle $ , which provides insights into how 
$A$ evolves with time.
Here, we assume that the correlators are even in the time.
This evolution is governed by the Heisenberg equation of motion:
\begin{equation}\label{te}
    \frac{dA}{dt} = i [H,A]
\end{equation}
Here, the commutator $[H,.]$ , known as the quantum Liouvillian operator $\mathcal{L}$, plays a crucial role. It's a Hermitian superoperator. The formal solution to the equation of motion is expressed as:
\begin{equation}
   A(t) = e^{i \mathcal{L}t} A(0).
\end{equation}
To implement the recursion method effectively, besides $H$ and 
$A$, we need to define an inner product for operators within the Hilbert space associated with 
$H$ and 
$A$. This choice influences the nature of the resulting dynamic correlation function.

The heart of the Liouvillian representation in the recursion method lies in the orthogonal expansion of the observable under examination:
\begin{equation}\label{ortho}
    A(t) = \sum _{k=0}^{\infty} \phi_k(t) A_k.
\end{equation}
For classical systems,
$A_k$
comprises an orthonormal set of functions in phase space. In contrast, for quantum systems, it constitutes an orthonormal set of operators. Regardless, these sets span a Hilbert space, typically of infinite dimensionality. The Liouvillian operator acts on the vectors 
$A_k$ within this space. The orthogonal expansion is executed in two successive steps.

\begin{itemize}
    \item Determine a particular orthogonal basis $A_k$ in the Hilbert space of the dynamical variables by applying the Gram-Schmidt procedure with the Liouvillian $\mathcal{L}$ as the generator of the new direction.
    \item  Insert the expansion \eqref{ortho} into the equation of motion to obtain a set of differential equations for the time-dependent coefficients $ \phi_k(t)$.
\end{itemize}

As the first step, we note that the general inner product between the vectors $A$ and $ i \mathcal{L} A$ for arbitrary $A$ vanishes
\begin{equation}
    \langle A , i \mathcal{L} A \rangle =0.
\end{equation}
This simplifies the Gram-Schmidt orthogonalization process and results in the subsequent set of recurrence relations for the vectors$ A_k$
\begin{equation}
    A_{k+1} = i \mathcal{L} A_k + \Delta_k A_{k-1}, ~~~~~ k=0,1,2,...
\end{equation}
\begin{equation}
    \Delta_k = \frac{\langle A_k, A_k \rangle}{ \langle A_{k-1}, A_{k-1 }\rangle}~~~~~ k=1,2,3,..
\end{equation}
with $ A_{-1} =0 $
and $ A_0 = A$. 
The sequence of numbers $ \Delta_k$ contains all the information for the reconstruction of the fluctuation function $ \langle A(t) , A(0) \rangle$.

In the second step, we plug in the orthogonal expansion \eqref{ortho} into the equation of motion. The differential operator acts on the $ \phi_k(t)$ and the Liouvillian acts on the $A_k$, which yields the following set of coupled linear differential equations for the function $\phi_k(t) $: 
\begin{equation}
    \frac{d \phi_k(t)}{dt} = \phi_{k-1} (t) - \Delta _{k+1} \phi_{k+1} (t), ~~~~~ k=0,1,2,...
\end{equation}
with $ \phi_{-1}\equiv 0$, $ \phi_k(0) = \delta_{k,0}$. 
Unlike the vectors $A_k$, the functions $ \phi_k(t)$ can not be determined recursively.

If our goal is to determine the fluctuation function of the dynamical variable $ A(t)$, then it is sufficient to know just one of the functions $ \phi_k(t)$. Follows directly from the orthogonal expansion 
\begin{equation}
    \phi_0(t) = \frac{\langle A(t), A(0) \rangle}{ \langle A(0) , A(0)\rangle}.
\end{equation}

There is a way to calculate the $\Delta_k $
 sequence for specific correlation functions of a given model system. It is called the moment expansion. The normalized fluctuation function can be expanded in a Taylor series
 \begin{equation}
     \phi_0 (t) = \sum_{k=0}^\infty \frac{i^{2k} t^{2k}}{(2k)!} M_{2k}
 \end{equation}
with $ M_0 \equiv 1$. 
The coefficients $ M_{2k}$ are the frequency moments of the normalized  spectral density
\begin{multline}
     M_{2k} = \langle \omega ^{2k}\rangle = \int _{-\infty} ^ \infty \frac{d\omega}{ 2 \pi} \omega ^{2k} f(\omega) \\ =i^{2k} \Big[ \frac{d^{2k}}{dt^{2k}} \phi_0(t)\Big]_{t=0}, ~~~~~ k=1,2,...
\end{multline}
while
\begin{equation}
    f(\omega) = \int _{-\infty} ^ \infty d\omega e^{i \omega t } \phi_0 (t).
\end{equation}
for a given set of moments $ M_{2k}, k=0,1,..., K$ with $ M_0 = 1$ the first $K$ coefficients $ \Delta_n$ are determined by 
\begin{equation}
    M_{2k}^{(n)} = \frac{M_{2k}^{(n-1)}}{\Delta_{n-1}} - \frac{M_{2k}^{(n-2)}}{\Delta_{n-2}},~~~ \Delta_n = M_{2n}^{(n)}
\end{equation}
for $ k=n, n+1, ... ,K$ and $ n = 1,2,...,K$, and with set values $ M_{2k} ^{(0)} = M_{2k}$,$M_{2k}^{-1} =0$, $ \Delta _{-1} = \Delta_0 = 1$.

The set of coefficients $ \Delta_n$ is equivalent to the square of the set of $b_n^2$ as discussed earlier.

\textbf{Recursion method of the quantum Hamiltonian system in its ground state:}

This application of the recursion method is tailored for investigating dynamic correlation functions within the quantum Hamiltonian system's ground state. An essential preliminary step in more practical scenarios involves identifying the ground state wave function of the system.

For a given quantum Hamiltonian $ H$ and its ground state wave function $ \ket{\phi_0}$, our goal is to determine the normalized correlation function of the dynamical variable represented by the Hermitian operator $A$
\begin{equation}
    C(t) = \frac{\bra{\phi_0} A(t) A(0) \ket{\phi_0}}{\bra{\phi_0} A(0) A(0) \ket{\phi_0}}
\end{equation}
In such a case the result of the Lanczos algorithm is two sets of coefficients $ a_k$ and $ b_k$. The relation between the moments and these sets of coefficients are most conveniently expressed in terms of two arrays of auxiliary quantities $ L_k ^{(n)} $ and  $ M_k ^{(n)} $:

Given a set of moments $ M_0 \equiv 1$, $ M_1,..., M_{2K+1}$ the coefficients $ a_0,..., a_K$ and $ b_1, ... ,b_K$ are obtained by initializing 
\begin{equation}
    M^{(0)} = (-1)^k M_k, ~~~~~ L_k^{(0)} = (-1)^{k+1} M_{k+1}
\end{equation}
for $ k=0,..., 2K$ and then applying the recursion relations \cite{iizuka2023krylov}
\begin{equation}\label{ab}
    \begin{split}
        M_k^{(n)}& = L_k^{(n-1)} - L_{n-1}^{(n-1)} \frac{ M_k^{(n-1)}}{M_{n-1}^{(n-1)}}
        \\
        L_k^{(n)} & = \frac{M_{k+1}^{n}}{M_n^{(n)}} - \frac{M_k^{(n-1)}}{M_{n-1}^{(n-1)}}
    \end{split}
\end{equation}
for $ k= n,..., 2K-n+1$ and $ n=1,..., 2K$. The resulting coefficients are
\begin{equation}\label{33333}
    b_n = \sqrt{ M_n^{(n)}}, ~~~~~ a_n = - L_n ^{(n)}, ~~~~~ n=0,...K.
\end{equation}

\section{Simple example: Krylov Complexity of free field theory}

For a single harmonic oscillator, the Euclidean two-point function must obey the equation 
\begin{equation}
    \big(- \frac{d^2}{d\tau ^2} + \omega ^2 \big) \langle X( \tau ) X(0) \rangle = \delta ( \tau). 
\end{equation}
The solution to this equation is 
\begin{equation}
     C_0 (\tau)= \langle X( \tau ) X(0) \rangle = \frac{1}{2 \omega} e ^{- \omega|\tau|}
\end{equation}
which can also be found by using the path integral method.
To find the finite-temperature two-point function, one can use the method of images 
\begin{equation}
    G_\beta (\tau) = \sum_{n = -\infty}^ \infty C_0 (\tau + n \beta ) .
\end{equation}
For simplicity, consider the case that $ 0< \tau < \beta $, then we have 
\begin{equation}
\begin{split}
    G_0 ( \tau)  &= \sum_{n= - \infty } ^ {-1}  \frac{1}{2\omega} e ^{\omega (\tau + n\beta )} + \sum _{n= 0}^\infty \frac{1}{2 \omega} e ^{-\omega (\tau + n\beta )}
    \\
    & = \frac{e ^ {\beta \omega - \tau \omega}}{2 \omega(-1 + \beta \omega)} +  \frac{e ^ {\tau \omega}}{2 \omega(-1 + \beta \omega)}.
\end{split}
\end{equation}
Therefore, the thermal correlator is given as 
\begin{multline}
      \Tr (e^{- \beta H } X(t) X(0) ) = G_\beta ( t)\\ = \frac{e ^ {\beta \omega - i t \omega}}{2 \omega(-1 + \beta \omega)} +  \frac{e ^ { it \omega}}{2 \omega(-1 + \beta \omega)}  
\end{multline}

In order to find the complexity we can use the inner product
which can be motivated or inspired by a two-sided correlator on the TFD state or KMS inner product as 
\begin{equation}\label{inner}
    \langle O_1, O_2\rangle = \Tr (e^{- \beta H /2} O_1^\dagger e^{- \beta H /2} O_2 )
\end{equation}
To find the inner product between the single harmonic oscillator and its time-shifted we can use the thermal two-point function and shift the time as $ t \rightarrow t - i \beta /2$
\begin{multline}
       \Tr (e^{- \beta H/2 } X(t) e^{- \beta H/2 }X(0) ) = G_\beta ( t- i \beta/2)\\ = \frac{e^{\beta \omega /2}}{2 \omega (-1 + e^{\beta \omega})} e^{it\omega} +\frac{e^{\beta \omega /2}}{2 \omega (-1 + e^{\beta \omega})} e^{-it\omega} 
\end{multline}

considering a free quantum field on a circle of length $L$. We can expand it in modes and get a collection of harmonic oscillators with frequency $ \omega_j$. In the following, we consider a J number of modes over the ground state and thermal states respectively.

\subsection{Krylov complexity of the operator $X$ over the ground state }

\begin{figure}
  \centering
  \begin{subfigure}[b]{0.7\linewidth}
    \includegraphics[width=\linewidth]{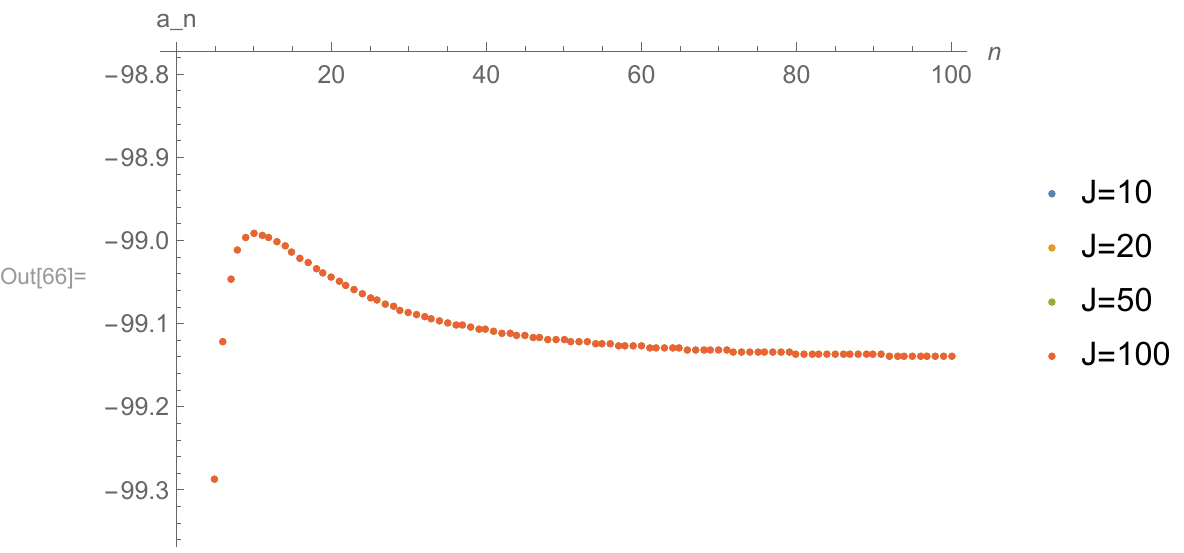}
    \caption{$a_n$}
  \end{subfigure}

  \begin{subfigure}[b]{0.7\linewidth}
    \includegraphics[width=\linewidth]{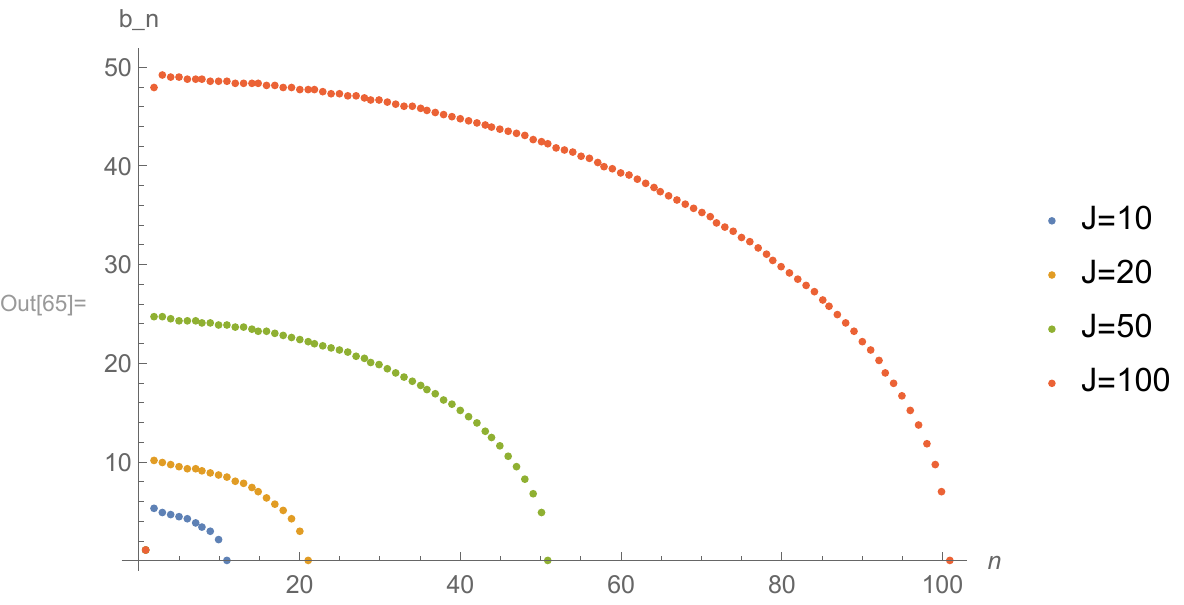}
    \caption{ $b_n$}
  \end{subfigure}

  \caption{The  non-zero values of $a_n$ and $b_n$ for different $J$. The plots of $a_n$ are on top of each other for different values of $J$, and only the number of nonzero values of $a_n$ will increase as one increases the $J$. }
  \label{overgs}
\end{figure}

The correlator for $J$ different modes of harmonic oscillator in the ground state is 
\begin{equation}
    C(t) = \frac{1}{N} \sum_{j=1}^J \frac{1}{2 \omega_j} e^{-i \omega_j t }
\end{equation}
while $N$ in the normalization factor such that $C(t=0) =1 $. The moments are 
\begin{equation}
    M_n = \frac{1}{N} \sum_{j=1}^J \frac{1}{2 \omega_j} \frac{(-i \omega_j)^n}{i^n}.
\end{equation}
Here both sets of odd and even moments are nonzero, thus we get the nonzero values for both sets of $a_n$ and $b_n$. In Fig. \ref{overgs}, one can find the non-zero value of  
 $a_n$ and $b_n$ for different value of $J$.




\subsection{Krylov complexity of the operator $X$ over the thermal state}

\begin{figure}
  \centering
  \begin{subfigure}[b]{0.7\linewidth}
    \includegraphics[width=\linewidth]{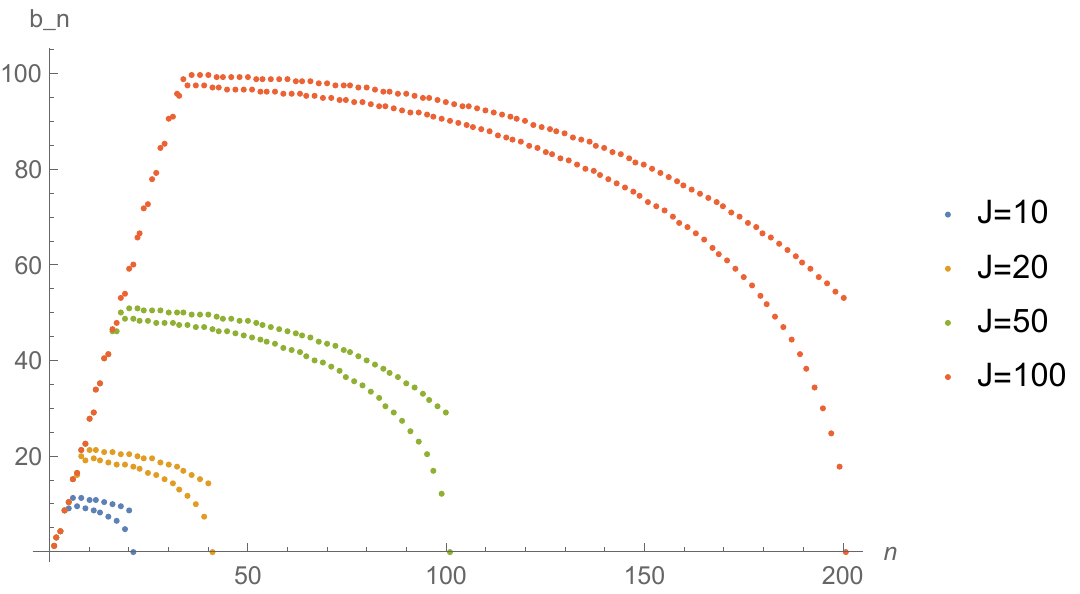}
    \caption{$\beta =1$}
  \end{subfigure}

  \begin{subfigure}[b]{0.7\linewidth}
    \includegraphics[width=\linewidth]{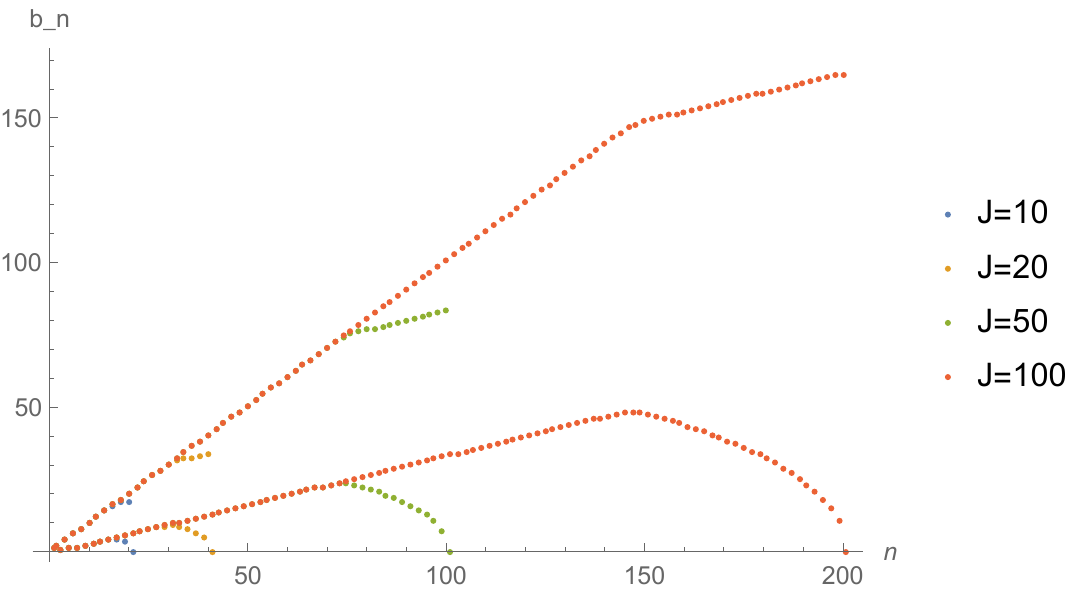}
    \caption{$\beta=5$}
  \end{subfigure}

  \caption{The non-zero $b_n$ for different value of $J$.}
  \label{fig1}
\end{figure}

\begin{figure}
  \centering
  \begin{subfigure}[b]{0.45\linewidth}
    \includegraphics[width=\linewidth]{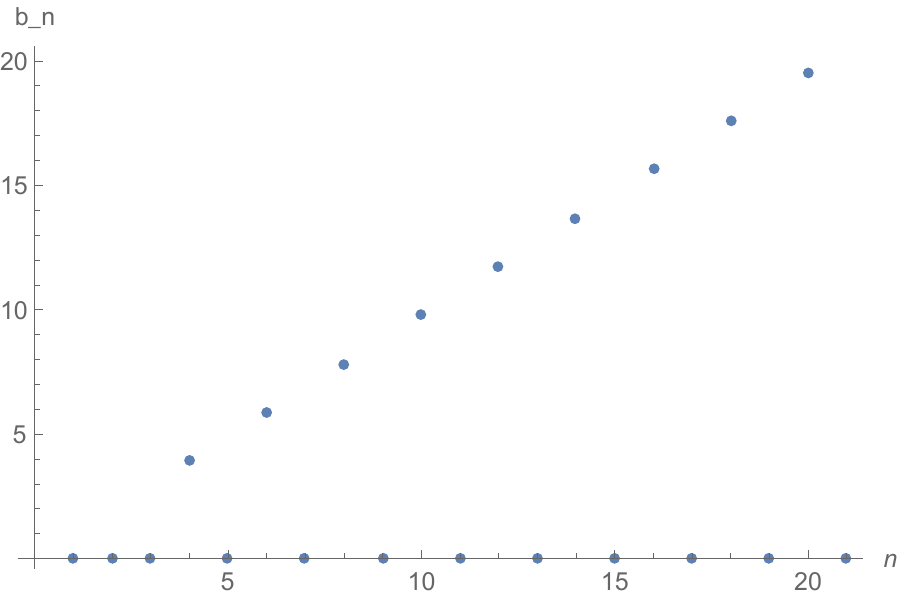}
    \caption{$J =10$}
  \end{subfigure}
  \begin{subfigure}[b]{0.45\linewidth}
    \includegraphics[width=\linewidth]{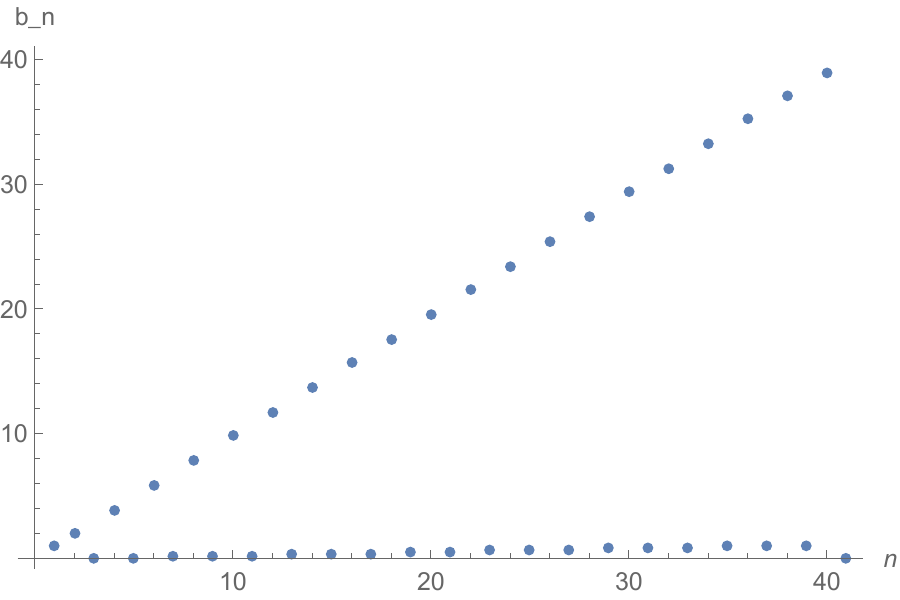}
    \caption{$J=20$}
  \end{subfigure}

   \begin{subfigure}[b]{0.45\linewidth}
    \includegraphics[width=\linewidth]{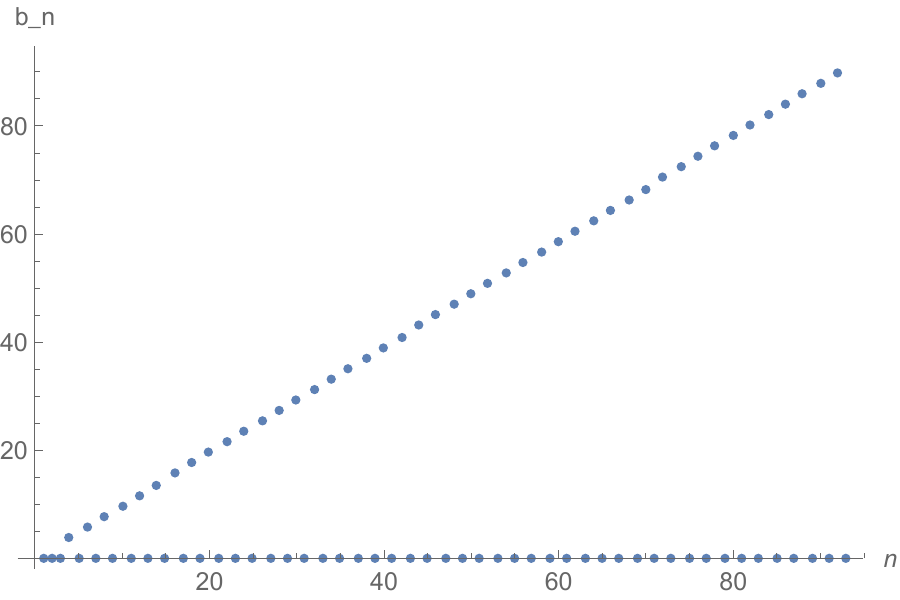}
    \caption{$J=50$}
  \end{subfigure}
  \begin{subfigure}[b]{0.45\linewidth}
    \includegraphics[width=\linewidth]{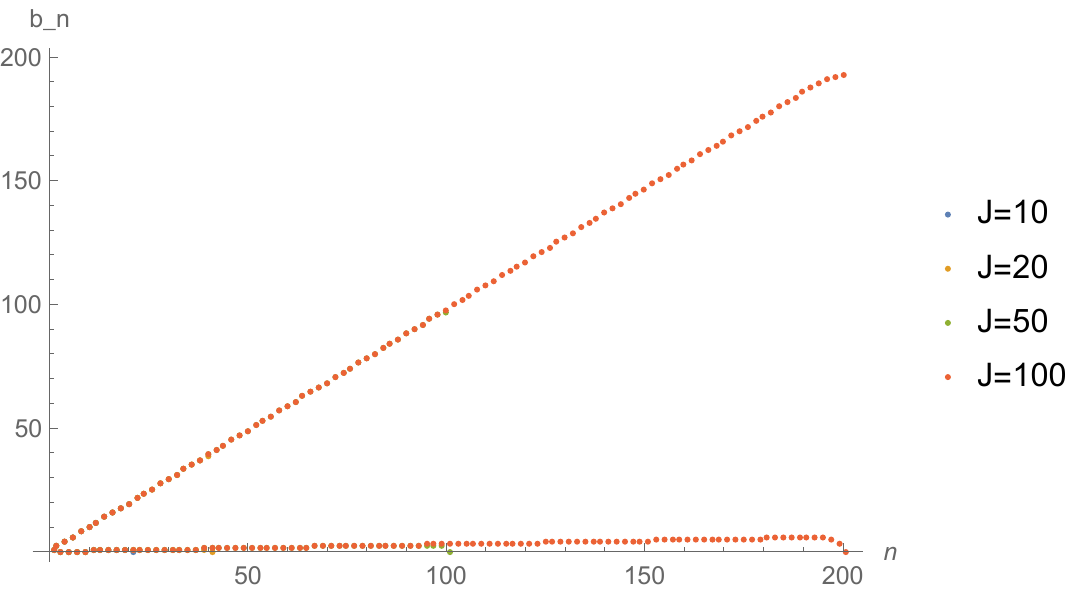}
    \caption{$J=100$}
  \end{subfigure}

  \caption{The non-zero $b_n$ for $ \beta =10$ and different value of $J$. They contain two linear branches with the same slopes but the numbers of non-zero value of $b_n$  depend on $J$ and it increases when $J$ increases. }
  \label{fig2}
  \end{figure}

To find the correlator we use the inner product defined in \eqref{inner}.
The correlator for $J$ different mode of the harmonic oscillator in the thermal state with inverse temperature $\beta $ is given by 
\begin{multline}
     C(t, \beta ) = \frac{1}{N} \sum _{j=1} ^J \frac{e^{\beta \omega_j /2}}{2 \omega_j (-1 + e^{\beta \omega_j})} e^{it\omega_j} \\+\frac{e^{\beta \omega_j /2}}{2 \omega_j (-1 + e^{\beta \omega_j})} e^{-it\omega_j}
\end{multline}
while 
\begin{equation}
    \omega_j = j \frac{\pi}{ 2 L}
\end{equation}
and $N$ the normalization factor such that 
$  C(t=0, \beta )=1$. The moments are 
\begin{equation}
    M_n =  \frac{1}{N} \sum _{j=1} ^J\frac{e^{\beta \omega_j /2}}{2 \omega_j (-1 + e^{\beta \omega_j})} \Big[\frac{(i \omega_j)^n}{i^n}+ \frac{(-i \omega_j)^n}{i^n}\Big].
\end{equation}
One can calculate the Lanczos coefficients using \eqref{33333}.
As it is clear $ M_{2n+1} = 0$ and thus
\begin{equation}
    a_n =0 ~~~~~~ \forall n.
\end{equation}
In Fig. \ref{fig1} and \ref{fig2}, one can see the behavior of the non-zero $b_n$ for different values of $J$ and $ \beta$.
In general, in this case, $b_n$ has two branches. For small $n$, it increases linearly, and at some point, it starts to decrease and goes to zero. The number of non-zero valued $b_n$ increases as we increase the $J$ and it is almost twice the value of $J$ for this range of $\beta$. Considering both positive and negative modes in the thermal case, the number of non-zero $b_n$ is equal to the number of different modes (in this case 2J).
As $ \beta$ increases, the linear behavior of the plots is dominant, and for $ \beta =10$ in Fig. \ref{fig2} one can see that we just have two linear branches. Moreover, by increasing the $ \beta$ the branches get more separated, and in the high value of $\beta $, it means the small value of $T$ the second branch is getting to vanish and we will reach the one linear branch as in the ground state. However, for a fixed $\beta$, the slopes of two branches remain constant. As one can see in Fig. \ref{fig1} the linear growth part of the plots for different $J$ are on top of each other.

\begin{figure}
  \centering
  \begin{subfigure}[b]{0.75\linewidth}
    \includegraphics[width=\linewidth]{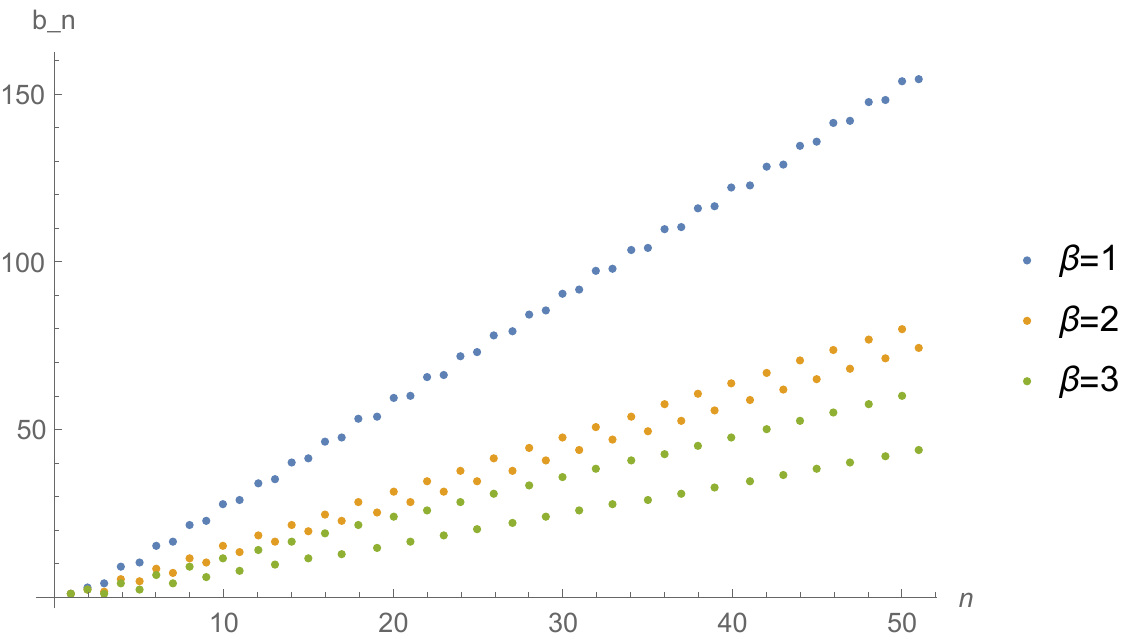}
    \caption{}
  \end{subfigure}
  
  \begin{subfigure}[b]{0.75\linewidth}
    \includegraphics[width=\linewidth]{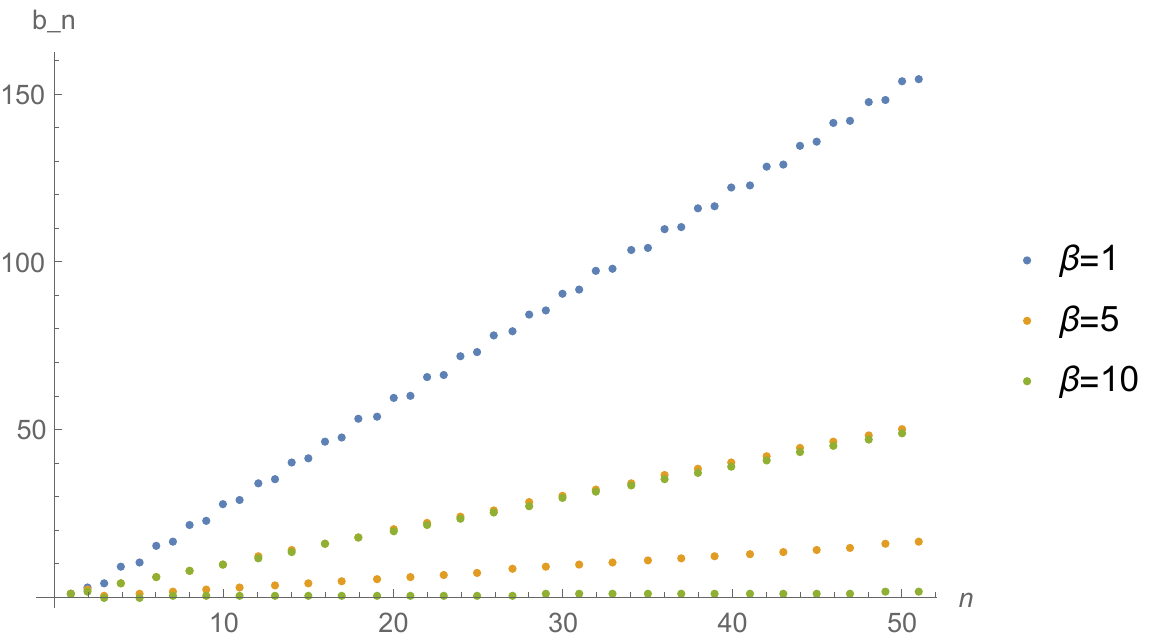}
    \caption{}
  \end{subfigure}

  \caption{The non-zero $b_n$ for the $ J $ goes to infinity limit.}
  \label{fig3}
\end{figure}

\begin{figure}

    \includegraphics[width=\linewidth]{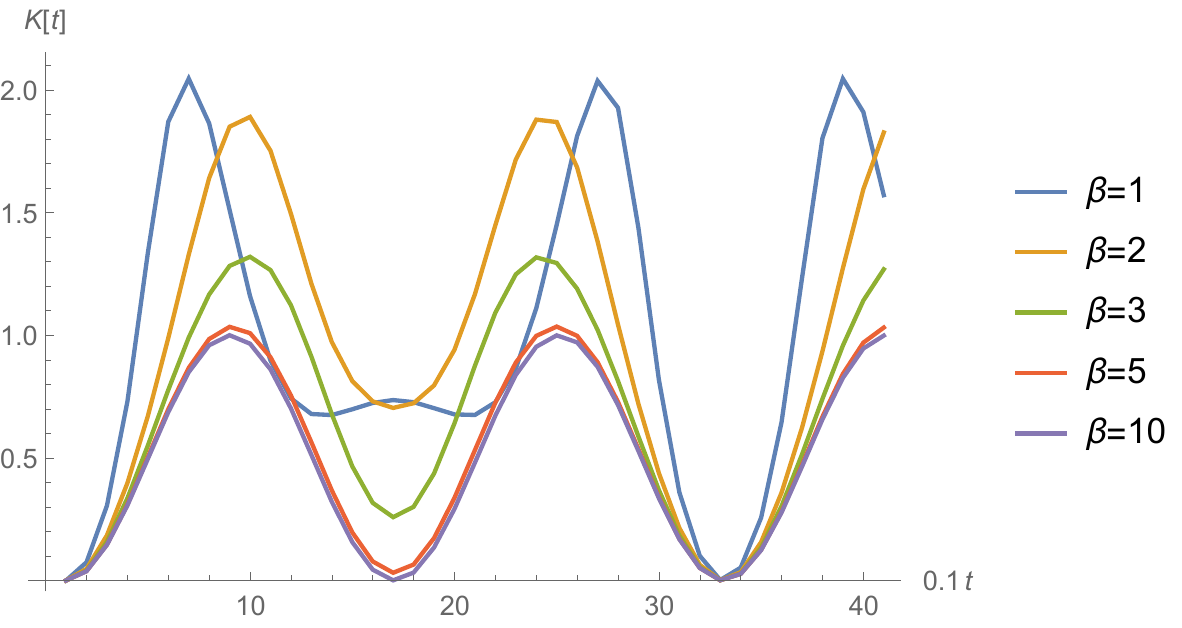}
    \caption{Krylov complexity as a function of time for different values of $\beta$.}

  \label{fig4}
\end{figure}

In Fig. \ref{fig3}, one can see the behavior of $b_n$ when $ J \rightarrow \infty$. It contains two linear branches and the slopes of two branches for different values of $\beta$ are different. Finally, in Fig. \ref{fig4}, one can see the behavior of the Krylov complexity for different values of $\beta$. As correlators are periodic in time, the Krylov complexity is also periodic with period of $4L$.  

In \cite{du2022krylov}, the authors discussed the simple example of harmonic oscillator analytically. In particular, they find that for a very generic choice for the frequency of the modes, only the first $2J$ Lancsoz coefficients are nonzero. For just one harmonic oscillator the theory describes with the Hamiltonian 
\begin{equation}
    H = \frac{1}{2} (X^2 + P^2)
\end{equation}
while $[X,P] = i\hbar $.
The position operator can be written in terms of the creation and annihilation operator 
$ X = \sqrt{\frac{\hbar}{2}} (a + a ^\dagger)$.
The calculation for the momentum operator is similar to the position operator. 
From the partition function, one can include the normalization factor 
\begin{equation}
    \tr (e^{-\beta H}) = \frac{1}{ 2 \sinh(\frac{\beta \hbar}{2})}. 
\end{equation}
we find 
\begin{equation}
    || X||^2 = || P||^2 = \frac{\hbar}{2 \sinh(\frac{\beta \hbar}{2})}.
\end{equation}
One can apply the Lanczos algorithm starting from a normalized operator 
\begin{equation}
    O_0 =\sqrt{\frac{2 \sinh(\beta \hbar / 2) }{\hbar}} X
\end{equation}
The first recursion gives 
\begin{equation}
    O_1 =- i \sqrt{\frac{2 \sinh(\beta \hbar / 2) }{\hbar}} P
\end{equation}
while 
\begin{equation}
    b_1 = \hbar
\end{equation}
Then the second operator in the recursion actually vanishes $ O_2 =0 , b_2=0$. 
So the harmonic oscillator is a rather trivial model with the Lanczos algorithm terminating at the second step. 

To generalize this case, consider a quantum system of $N$ decoupled harmonic oscillators of different frequencies 
\begin{equation}
    H = \sum _{i=0}^{J} \frac{1}{2} ( P_i^2 + \omega_i ^2 X_i ^2)
\end{equation}
with a properly normalized initial operator 
\begin{equation}
    O_0= \sum _{i=1}^{J} X_i.
\end{equation}
It is easy to compute the moments in the case 
\begin{equation}
    M_n = \frac{1}{N} (\hbar ^n \sum_{i=1} ^J \omega _i ^n), \qquad n: even
\end{equation}
The determinant in 
\begin{equation}
    b_1^{2n} b_2^{2(n-1)}...b_n^2 = \det(M_{(i+j)})_{0\leq i, j \leq n}
\end{equation}
vanishes for 
\begin{equation}
    n \geq 2J,
\end{equation}
so the Lanczos algorithm terminates at the $2J$ steps with $b_{2J+1}=0$.

For a more general discussion on free theory, one can look at \cite{avdoshkin2022krylov}.
They consider free massive scalar and Dirac fermion in $d$ spacetime dimension. 
In the first case, Lancsoz coefficients split into even and odd branches, growing linearly with $n$ albeit with different intercepts. $b_n$ grows linearly with the universal slope, but even and odd branches have different finite terms.  
In the second case of free massless fermions $C(t)$ is not an even function, hence besides $b_n$, Lancsoz coefficients also include $a_n$. In \cite{avdoshkin2022krylov}, one can see the numerical results as a function of $\beta$. 

They also consider a CFT on a sphere and calculate Lancsoz coefficients and Krylov complexity associated with the thermal two-point function of the model. 
They consider $4 d$ free massless scalar compacted on a $S^3$. The corresponding two-point function has some singularity on the imaginary time axis. 
The correlator is in terms of a parameter $R$ which is the radius of $S^3$ measured in the units of $\beta$ which is the radius of $S^1$. Their numerical results are good for $ R <1$. The Lanczos coefficients split into even and odd branches which grow linearly with $n$ but with different slopes. The same as our results in the thermal case. The behavior of Krylov complexity is the same as the $\beta \sim 1$ of our results (see Fig. \ref{fig4}).

\section{Matrix Quantum Mechanics}

This chapter is based on \cite{das1992one, das1990string}.
The one dimension takes to be timelike and the Lagrangian defines the theory
\begin{equation}
    \mathcal{L} = \Tr \Big( \frac{1}{2} \dot{M}^2(t) - V(M)\Big) 
\end{equation}
where $M$ is a Hermitian matrix variable. The Lagrangian is invariant under a global $U(N)$ symmetry, $ M \rightarrow U ^{-1} M U$ with the conserved $U(N)$ angular momentum
\begin{equation}
    J = i [ M, \dot{M}]\qquad \frac{dJ}{dt}=0.
\end{equation}
The quantum theory then is  defined by the Hamiltonian 
\begin{equation}
    H = \Tr \Big( -\frac{1}{2} \frac{\partial^2}{\partial M^2}  + V(M)\Big)
\end{equation}
and we restrict ourselves to the singlet sector $ J \ket{~} =0 $.

A set of basic singlet vertex operators is given by 
\begin{equation}
    \phi _m = \Tr (M^m). 
\end{equation}
To work with the collective field theory approach, a natural set of singlet operators is given by the vertex operators
\begin{equation}
    \phi_k = \Tr (e^{ik M})
\end{equation}
and one considers the collective field as its Fourier transform
\begin{equation}
    \phi(x) = \int \frac{dk}{2\pi} e ^{-ikx} \phi _k = \int \frac{dk}{2\pi} e ^{-ikx}\Tr (e^{ik M}).  
\end{equation}
In terms of the eigenvalues 
\begin{equation}
    M = U^{-1} diag ( \lambda _i) U 
\end{equation}
one has $ \phi_k = \sum _{i=1} ^N e^{ik \lambda_i} $
and thus 
\begin{equation}
    \phi(x) = \sum _{i=1} ^N \delta(x-\lambda_i).
\end{equation}
$ \phi(x)$ is simply the density of eigenvalues $ \lambda_i$.
The Collective field is constrained by 
\begin{equation}\label{cons}
    \phi(x) \geq 0, \qquad \int \phi(x) dx = N 
\end{equation}
and other constrained which disappear as $ N \rightarrow \infty$.

To reformulate the theory with $ \phi $ as the coordinate, one not only needs to change variables in the Hamiltonian but also to rescale the wavefunctions by the Jacobian of the transformation from $M$ to $ \phi$. While the Jacobian is singular for finite $N$, the $ N \rightarrow \infty $ may be found from the hermiticity of the Hamiltonian.  One can compute 
\begin{equation}
    \begin{split}
        \omega &(k, \phi) = -\frac{\partial^2}{\partial M ^2} \phi_k = k^2 \int _{0} ^1 d \alpha \phi _{\alpha k } \phi _{k(1-\alpha)}
        \\
        \Omega & (k,k'; \phi)= \frac{\partial \phi_k}{\partial M } \frac{\partial \phi_k'}{\partial M }= kk' \phi _{k + k'}
    \end{split}
\end{equation}
One can easily verify the following useful identity
\begin{equation}
    \omega (k, \phi) = \int dk' \Omega (k,k',\phi) \frac{1}{|k'|} \phi _{-k'}
\end{equation}
The Fourier transform of $ \omega(k,\phi)$ is the singular form
\begin{equation}
    \omega (x, \phi) = 2 \partial_x \int \frac{\phi(x) \phi (y) }{x-y} dy 
\end{equation}
In the end, one can write down the following field theory Hamiltonian
\begin{multline}
           H_\phi = \int dx~ \Big( \frac{1}{2} \partial_x \Pi \phi \partial_x \Pi + V(\phi) \phi(x) - \mu_F \big(\phi(x) \\- \frac{N}{V}\big)  
        + \frac{1}{2} \int dx~ \phi(x) \big(\int dy \frac{\phi(y)}{x-y}\big)^2 \Big)
\end{multline}
where $ \Pi $ is the momentum conjugation to $ \phi$, $ -i \frac{\delta}{\delta \phi(x)}$ and $ \mu _F$ represent a multiplier for the density constraint and we also have some additional singular terms associated with the derivative terms. 
The kinetic energy piece is local.
The effective potential is given by 
\begin{equation}
    V_{eff} = \frac{1}{2} \int dx \phi(x) \big(\int dy \frac{\phi(y)}{x-y}\big)^2 - \int (\mu_F -V(x)) \phi(x) ~dx
\end{equation}
One can evaluate the integral and find 
\begin{equation}
    V_{eff}= \int dx~ \Big( \frac{\pi ^2}{6} \phi ^3 (x) - \big(\mu_F -V(x)\big) \phi(x)\Big)
\end{equation}
We also have two other terms which are of lower order 
\begin{multline}
    \Delta V = \frac{1}{2} \int_{y=x} dx~ \phi(x) \partial_x \partial_y \ln (x-y) \\+ \frac{1}{2} \int \frac{\partial \Omega}{\partial \phi} \int \ln |x-y| \phi (y). 
\end{multline}
They do not contribute to the planar limit but begin to contribute in the first torus correction.  

We should find the classical equation of motion. Since the constraints \eqref{cons} should satisfy, the ground state has $ \partial \Pi / \partial x =0 $ and in the leading order minimize 
\begin{equation}
 V(\phi) - \mu_F \int \phi(x) ~dx.   
\end{equation}
This gives
\begin{equation}
    \phi_0(x) = \Bigg\{
    \begin{matrix}
        \frac{1}{\pi} \sqrt{2 (\mu_F - v(x))} \qquad\qquad~~~ |x| < \Lambda
        \\
      0 \qquad\qquad\qquad \qquad \qquad\qquad|x|> \Lambda
    \end{matrix}
\end{equation}
where $ \Lambda$ is the point at which the square root vanishes. the planar ground state energy is then given by 
\begin{equation}
    E_{0, GS} = \mu_F - \frac{1}{3 \pi} \int dx~ \big(2 (\mu_F - v(x))\big)^{3/2}.
\end{equation}

We now proceed to the computation of the propagator. This corresponds to the study of fluctuations in the collective field method. By shifting the field 
\begin{equation}
    \phi (x,t) = \phi_0(x) + \xi (x,t)
\end{equation}
the propagator is determined by the quadratic action 
\begin{equation}
  S= \int dx dt~ \Big( \frac{1}{2} \partial^{-1}_x \date{\xi} \frac{1}{\phi_0(x)} \partial^{-1}_x \date{\xi} + \frac{1}{2} \pi ^2  \phi_0(x) \xi ^2\Big).  
\end{equation}
It is convenient to introduce a new variable $q$ as
\begin{equation}
    q = \frac{1}{\pi} \int^x \frac{dx}{\phi_0(x)}.
\end{equation}
For a classical particle moving in the potential $ v(x)$, $q$ is the time taken for the particle to go from the origin to the point $x$. The range of $q$ is given by $ -L< q< L$ where $4L$ is the time period of the classical motion and it is determined by
\begin{equation}
    \frac{1}{\pi}  \int _0 ^ \Lambda \frac{dx}{\phi_0(x)} = L
\end{equation}
where $ \pm \Lambda$ are the turning points of the classical motion.
by redefining the field variable
\begin{equation}
    \xi = \frac{1}{\phi_0(x)} \eta
\end{equation}
we will give
\begin{equation}
    S = \pi ^3 \int dt \int _{- L} ^{L} dq \big( \frac{1}{2} \partial_q^{-1} \dot{\eta} \partial_q^{-1} \dot{\eta}- \frac{1}{2} \eta ^2 \big). 
\end{equation}
Notice that the background field $ \phi _0(x)$ has disappeared. The only remnant is the new integration region $ (-L, L)$ for the variable $q$. With the further transformation
\begin{equation}
     \eta = \partial_q \psi
\end{equation}
the action is brought into the form 
\begin{equation}
    S = \pi ^3  \int dt \int _{- L} ^{L} dq \Big( \frac{1}{2} (\partial_t \psi)^2 - \frac{1}{2} (\partial_q \psi)^2 \Big) 
\end{equation}
the propagator of the scalar field $ \psi (q,t)$
are obtained by implementing the constraint 
\begin{equation}
    \frac{d}{dt} \int dx~ \phi (x) =0
\end{equation}
which leads to the Dirichlet boundary condition on $ \psi : \psi (-L,t) = \psi (L,t)=0$. The small fluctuation eigenfunctions are  found to be
\begin{equation}
    \psi_j(q) = \begin{cases}\frac{1}{\sqrt{L}} \sin (\frac{j \pi q}{L}) \qquad\qquad j=0, 1 ,2,...
    \\
    \frac{1}{\sqrt{L}} \cos ((j+\frac{1}{2})\frac{\pi q}{L})
    \end{cases}
\end{equation}
with the frequencies
\begin{equation}
    \omega_j = \frac{j\pi}{2 L} = j \omega_c\qquad\qquad j=0, 1 ,2,... ~. 
\end{equation}
The propagator is then 
\begin{equation}\label{propa}
    D(t-t'; q,q') = \int \frac{dE}{\pi} e^{i E(t-t')} \sum _j \frac{ \psi_j(q)  \psi_j(q')}{E^2 - \omega_j^2 +i\epsilon}.
\end{equation}

To find the two-point function in the matrix model, we have 
\begin{equation}
    \Tr M^n = (-i)^n \frac{\partial^n \phi_k}{\partial k^n} \Big|_{k=0}.
\end{equation}
In terms of the collective field, one can find that 
\begin{equation}
    \Tr M^n(t) =  \int dx ~x^n \phi(x,t),
\end{equation}
therefore
\begin{equation}
    \langle \Tr M^n(t) \Tr M^m(0) \rangle = \int dx dx'~ x^n x'^m \langle \phi(x,t) \phi(x',0) \rangle.
\end{equation}
By substituting 
\begin{equation}
    \phi(x,t) = \phi_0(x) + \frac{1}{\phi_0(x)} \partial_q \psi (x,t)= \phi_0(x) +\partial_x \psi(x,t) 
\end{equation}
we reach to 
\begin{equation}
    \langle \phi(x,t) \phi(x',0) \rangle = \phi_0 (x) \phi_0(x') +  \partial_x \partial_{x'}\langle \psi(x,t) \psi(x',0) \rangle
\end{equation}
and thus
\begin{equation}
\begin{split}
      \langle \Tr M^n(t) \Tr M^m(0) \rangle =  \int dx dx'&~ x^n x'^m ~\phi_0 (x) \phi_0(x') + 
      \\
       \int dx dx'~ x^n x'^m~&\partial_x \partial_{x'} \langle \psi(x,t) \psi(x',0) \rangle
\end{split} 
\end{equation}
and the connected two-point function in terms of the propagator in \eqref{propa} can be written as
\begin{multline}
       \langle \Tr M^n(t) \Tr M^m(0) \rangle_c =  \int dq dq'~ x^n[q] x'^m[q'] \\ ~\partial_q \partial_{q'} D (t; q,q'). 
\end{multline}

To evaluate the integration over $E$ in \eqref{propa} for $ t>0$ we can take the integration over upper half plane and in the case of $ t<0$ over the lower half plane and we will find that 
\begin{multline}
      D(t;q,q') = \theta(t) \sum_j \frac{\psi_j(q) \psi_j(q') e^{-i\omega_j t }}{i\omega_j } + \\\theta(-t) \sum_j \frac{\psi_j(q) \psi_j(q') e^{i\omega_j t }}{i\omega_j } 
\end{multline}
From now on we assume that $ t>0$ and thus
\begin{multline}
       D(t;q,q') = \sum_j \frac{e^{-i\omega_j t }}{i\omega_j L } \Big(\sin(\frac{j \pi q}{L}) \sin(\frac{j \pi q'}{L}) \\+ \cos ((j+\frac{1}{2})\frac{\pi q}{L}) \cos ((j+\frac{1}{2})\frac{\pi q'}{L})\Big)
\end{multline}

therefore
\begin{widetext}
\begin{equation}
    \begin{split}
       \langle \Tr M^n(t)& \Tr M^m(0) \rangle_c 
       \\
       =& \int dq dq'~ x^n[q] x'^m[q']~ \Big( \sum_j \frac{e^{-i\omega_c j t } j \pi^2}{i\omega_c L^3 } \cos(\frac{j \pi q}{L}) \cos(\frac{j \pi q'}{L})
       \\
       & + \sum_j \frac{e^{-i\omega_c (j + 1/2)^2 t } j \pi^2}{i\omega_c j L^3 } \sin((j+1/2)\frac{ \pi q}{L}) \sin((j+1/2)\frac{ \pi q'}{L})\Big)
        \\
       =& \sum_j \frac{e^{-i\omega_c j t } j \pi^2}{i\omega_c L^3 } \int dq x^n[q] \cos(\frac{j \pi q}{L}) \int dq' x'^n[q'] \cos(\frac{j \pi q'}{L})
       \\
       &+ \sum_j \frac{e^{-i\omega_c j t } (j+1/2)^2 \pi^2}{i\omega_c j L^3 } \int dq x^n[q] \sin((j+1/2)\frac{ \pi q}{L}) \int dq' x'^n[q'] \cos((j+1/2)\frac{\pi q'}{L})
    \end{split}
\end{equation}
\end{widetext}

\subsection{Quadratic Potential}

We start with the free theory. Taking $ v(x) = x^2$, we have
\begin{equation}
    \phi_0(x) = \frac{1}{\pi} \sqrt{2 \mu_F - 2x^2},
\end{equation}
thus the $x$ variable in terms of $q$ can be find as 
\begin{equation}
    x[q] = \sqrt{mu_F} \sin( \sqrt{2}q).
\end{equation}
We have $ - \frac{\pi}{2} < q< \frac{\pi}{2}$ and so 
\begin{equation}
    L = \frac{\pi}{2\sqrt{2}}.
\end{equation}
In the end, for free theory, we find that 
\begin{equation}
\begin{split}
      \langle \Tr M^n(t) \Tr M^m(0) \rangle_c&= 
      \\
       \frac{16}{\pi} (\sqrt{\mu_F})^{m+n} \int dq~ &\sin ^m (\sqrt{2} q) \cos (2 \sqrt{2} j q )
       \\
       \int dq'~ \sin ^n &(\sqrt{2} q') \cos (2 \sqrt{2} j q')
\end{split}
\end{equation}
For some value of $ m,~n$, the result is as below
\begin{widetext}
\begin{equation}
\begin{split}
    \langle \Tr M^2(t) \Tr M^2(0) \rangle_c=& \frac{1}{2} \mu_F ^2 \pi e^{-i \sqrt{2}t}
    \\
    \langle \Tr M^4(t) \Tr M^4(0) \rangle_c=& \frac{1}{2} \mu_F ^4 \pi e^{-i \sqrt{2}t} + \frac{1}{16} \mu_F ^4 \pi e^{-i 2\sqrt{2}t}
    \\
    \langle \Tr M^6(t) \Tr M^6(0) \rangle_c=& \frac{225}{512} \mu_F ^6 \pi e^{-i \sqrt{2}t} + \frac{9}{64} \mu_F ^6 \pi e^{-i 2\sqrt{2}t} + \frac{3}{512} \mu_F ^6 \pi e^{-i 3\sqrt{2}t} 
\end{split}
\end{equation}
\end{widetext}

\subsection{Quatric Potential
}

Now let us consider the interacting theory, the simplest potential is 
\begin{equation}
    v(x)= x^2 + g x^4
\end{equation}
and we set $ 2 \mu_F = 1$ here. Hence, we have 
\begin{equation}
    q = \int \frac{dx}{\sqrt{1-2x^2 -2 g x^4}}
\end{equation}
by change of variable 
$ t = \sqrt{2 g }x / \sqrt{-1 + \sqrt{1+2 g}}$ we will reach to 
\begin{equation}
\begin{split}
     q =& \frac{ \sqrt{-1 + \sqrt{1 + 2 g }}}{\sqrt{2 g }} \int _0^t \frac{dt'}{\sqrt{(1-t'^2)(1+ \frac{1+g -\sqrt{1+2 g }}{g} t'^2)}}
     \\
    = &\frac{ \sqrt{-1 + \sqrt{1 + 2 g }}}{\sqrt{2 g }} F (t, - \frac{1+g -\sqrt{1+2 g }}{g})
    \\
       = &\frac{ \sqrt{-1 + \sqrt{1 + 2 g }}}{\sqrt{2 g }} F (  \frac{\sqrt{2 g} x}{ \sqrt{-1 + \sqrt{1_2 g }}}, - \frac{1+g -\sqrt{1+2 g }}{g})
\end{split}
\end{equation}
where 
\begin{equation}
    F(x,m) = \int_0 ^x \frac{dt}{\sqrt{(1-t^2)(1-mt^2)}}
\end{equation}
is the elliptic integral of the first kind.
The turning point of the classical particle is at $ \Lambda = \frac{1}{\sqrt{2 g}\sqrt{-1+\sqrt{1+2 g }}}$. Therefore
\begin{widetext}
\begin{equation}
    L = \frac{ \sqrt{-1 + \sqrt{1 + 2 g }}}{\sqrt{ 2g }} \int _0^1 \frac{dt}{\sqrt{(1-t^2)(1+ \frac{1 + g -\sqrt{1 + 2 g }}{g} t^2)}} =  \frac{ \sqrt{-1 + \sqrt{1 + 2 g }}}{\sqrt{2 g}} K (\frac{-(g + 1) +\sqrt{1 + 2 g }}{g})
\end{equation}
\end{widetext}
where 
\begin{equation}
    K(m)= \int_0 ^1 \frac{dt}{\sqrt{(1-t^2)(1-mt^2)}}
\end{equation}
is the complete elliptic integral of the first kind. Solving $x$ in terms of $q$, one can find that 
\begin{multline}
     x[q] = \frac{ 1}{\sqrt{ 2 g}} \sqrt{-1 + \sqrt{ 1 + 2g }} ~\\\textbf{sn} \Big( \frac{ 1}{ \sqrt{ 2g }} \sqrt{-1 + \sqrt{ 1 + 2 g}} (1+\sqrt{ 1 + 2 g}) ~q | \frac{1}{g}(-( 1 +g ) + \sqrt{ 1+ 2g }) \Big)   
\end{multline}

where $ \textbf{sn}(z|m)$ is the Jacobi elliptic function.

In order to find the connected two-point function, we need to calculate 
$ \int dq x^m \cos(j\pi q /L)$. To proceed, we can use the series definition of The Jacobi elliptic function 
\begin{equation}
    \textbf{sn}(z|m) = \frac{2\pi }{ \sqrt{m} K(m)} \sum _{n=0}^{\infty} \frac{q(m)^{n+1/2}}{1-q(m)^{2n+1}} \sin ((2n+1) \frac{\pi z}{2 K(m)})
\end{equation}
In our case 
\begin{equation}
     m = \frac{1}{g} (-(1 + g)+ \sqrt{1 + 2 g}) ,
\end{equation}
and
$ L / K(m) = \frac{1}{\sqrt{2 g}} \sqrt{-1+\sqrt{1 + 2 g}}$.
Let us first calculate the two-point function for the singlet $\Tr M^2 $. Thus, we have
\begin{widetext}
\begin{equation}
\begin{split}
     \int dq x^2[q] \cos(j\pi q /L)=
    - \frac{1}{2L} \sum _{n,l=0}^ \infty &\frac{q(m) ^ {n+l+1}}{(1- q(m)^{2n+1})(1- q(m)^{2l+1})}
    \\
    \int _{-L}^L dq &\sin((2n+1) \frac{\pi q }{2 L}) \sin((2l+1) \frac{\pi q }{2 L}) \cos(\frac{j \pi q}{L}).
\end{split}
\end{equation}
\end{widetext}
We have 
\begin{widetext}
\begin{equation}
\begin{split}
       \int _{-L}^L & dq\sin((2n+1) \frac{\pi q }{2 L}) \sin((2l+1) \frac{\pi q }{2 L}) \cos(\frac{j \pi q}{L})
       =
       \\
      & \frac{L}{2 \pi} \Big\{ \frac{\sin (j+l-n)\pi}{j+l-n} + \frac{\sin (j-l+n)\pi}{j-l+n} - \frac{\sin (1-j+l+n)\pi}{1-j+l+n} - \frac{\sin (1+j+l+n)\pi}{1+j+l+n}\Big\}
\end{split}
\end{equation}
\end{widetext}
and in the end 
\begin{equation}
    \int_{-L}^L dq~ x^2[q] \cos(\frac{j\pi q}{L}) = -\frac{1}{4}\{2A_j + B_j\}
\end{equation}
while
\begin{equation}
    \begin{split}
        A_j= & \sum _{l=0}^\infty \frac{q(m)^ {2l + 1+j}}{(1-q(m)^{2l+2j+1})(1- q(m)^{2l+1})}
        \\
        B_j=& \sum _{l=0}^{j-1} \frac{q(m)^ {j}}{(1-q(m)^{2l+1})(1- q(m)^{2j-2l-1})}.
    \end{split}
\end{equation}
Finally, we reach to
\begin{equation}
    \langle \Tr M^2(t) \Tr M^2(0) \rangle_c= \sum _{j=1}^\infty \frac{-i j \pi }{8 L^2} e ^{\frac{-i\pi j t }{2L}}  \{2A_j + B_j\}^2
\end{equation}

\section{Krylov complexity for 1-MQM via the Lanczos algorithm}

Now it is time to attempt to find the notion of Krylov complexity for the 1-MQM.

\subsection{Over the Ground State}

\begin{figure}
  \centering
  \begin{subfigure}[b]{0.75\linewidth}
    \includegraphics[width=\linewidth]{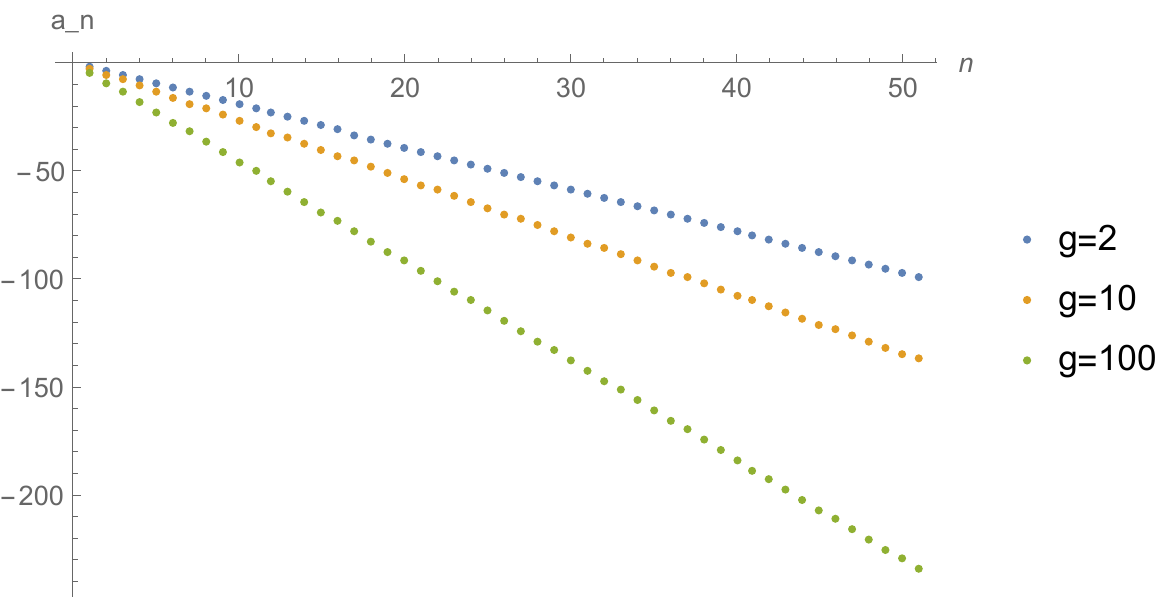}
    \caption{$a_n$}
  \end{subfigure}
  
  \begin{subfigure}[b]{0.75\linewidth}
    \includegraphics[width=\linewidth]{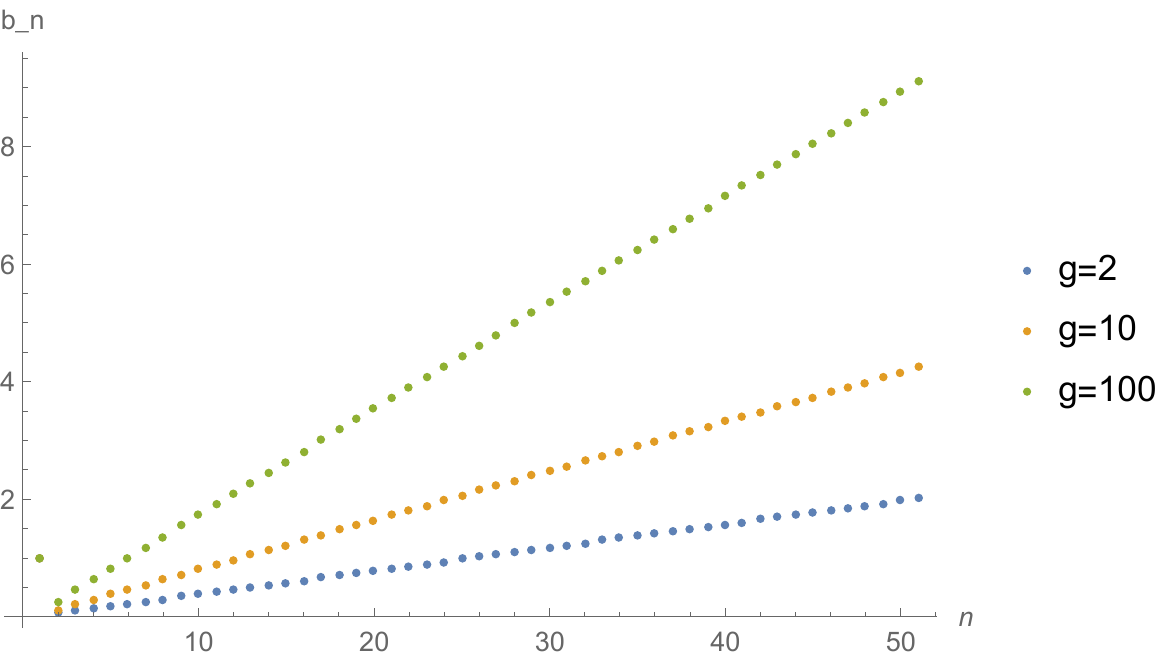}
    \caption{$b_n$}
  \end{subfigure}
  
   \begin{subfigure}[b]{0.75\linewidth}
    \includegraphics[width=\linewidth]{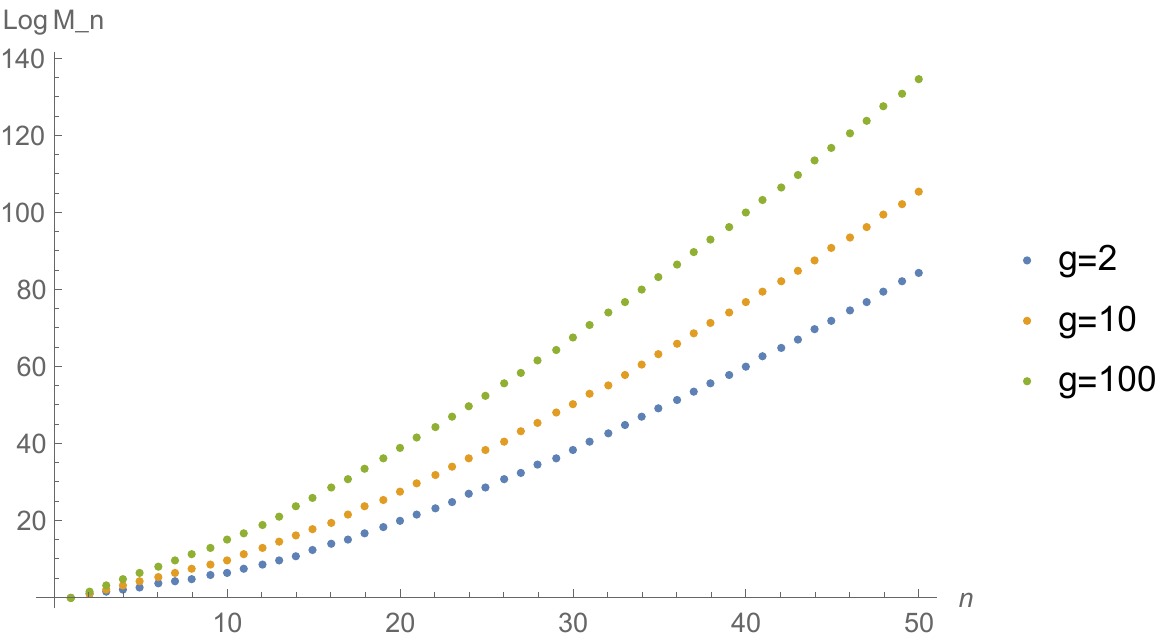}
    \caption{$\log M_n$}
  \end{subfigure}

  \caption{The moments and Lanczos coefficients for different values of $g$.}
  \label{fig7}
\end{figure}

The correlator in the ground state is  
\begin{equation}
    C(t) = \frac{1}{N} \sum_j \frac{-ij \pi }{8 L^2} e^{-ij \pi t /2 L} \{2A_j + B_j\}^2
\end{equation}
while $N$ is the normalization factor such that $ C(t=0)=1$. The moments are given by
\begin{equation}
    M_n = \frac{1}{N} \sum \frac{-i j \pi }{ 8 L^2} \{2A_j + B_j\}^2 (\frac{-ij \pi}{ 2L})^n \frac{1}{i^n}.
\end{equation}
In Fig. \ref{fig7}, one can see the moments and Lanczos coefficients of the 1-MQM  in the ground state for different values of $g$. 
The Lanczos coefficients have a linear behavior in that the absolute value of the slope increases for higher values of the $ g$ parameter. 
The slopes of the $a_n$ coefficients are negative while in the case of $b_n$, they are positive.

\begin{figure}

    \includegraphics[width=\linewidth]{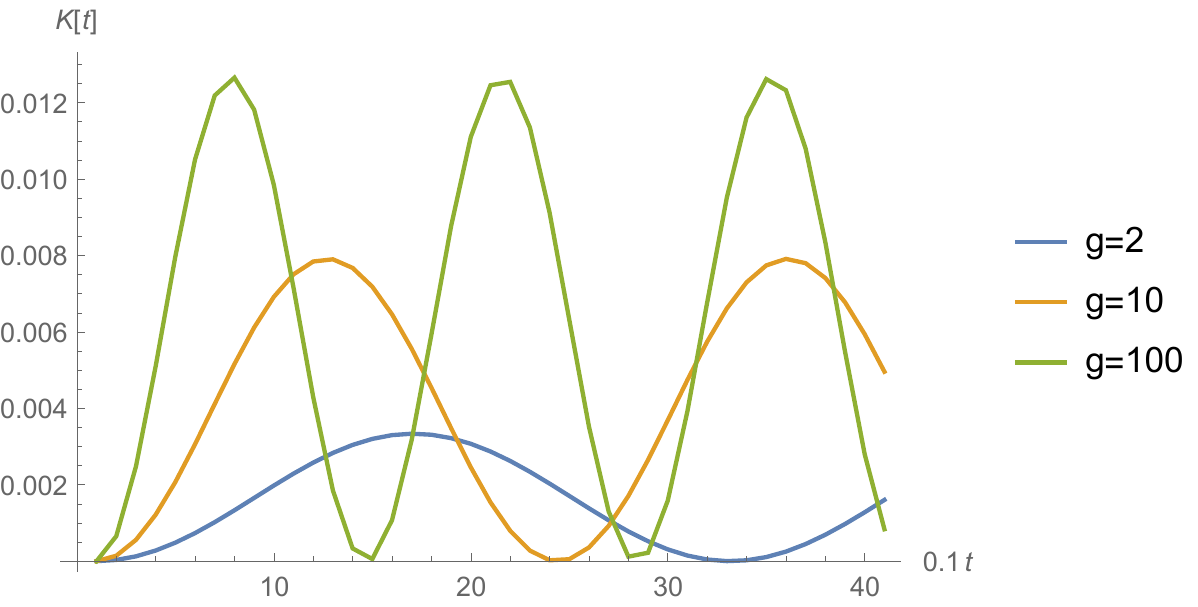}
    \caption{Krylov complexity of the matrix quantum mechanics in the ground state as a function of time for different values of $g$.}

  \label{fig8}
\end{figure}

Finally, in Fig. \ref{fig8}, one can find the Krylov complexity for different values of $g$. 
The peak of the complexity grows while g increases. The complexity is periodic as the correlation function is periodic. However, we should consider the behavior of the complexity as a function of time for the time less than the radius of convergence in the time direction. 
The period of complexity is related to the $L$ and it decreases while $g$ increases and it is expected that it saturates for infinite $g$.

\subsection{Over the Thermal State}

\begin{figure}
  \centering
  \begin{subfigure}[b]{0.70\linewidth}
    \includegraphics[width=\linewidth]{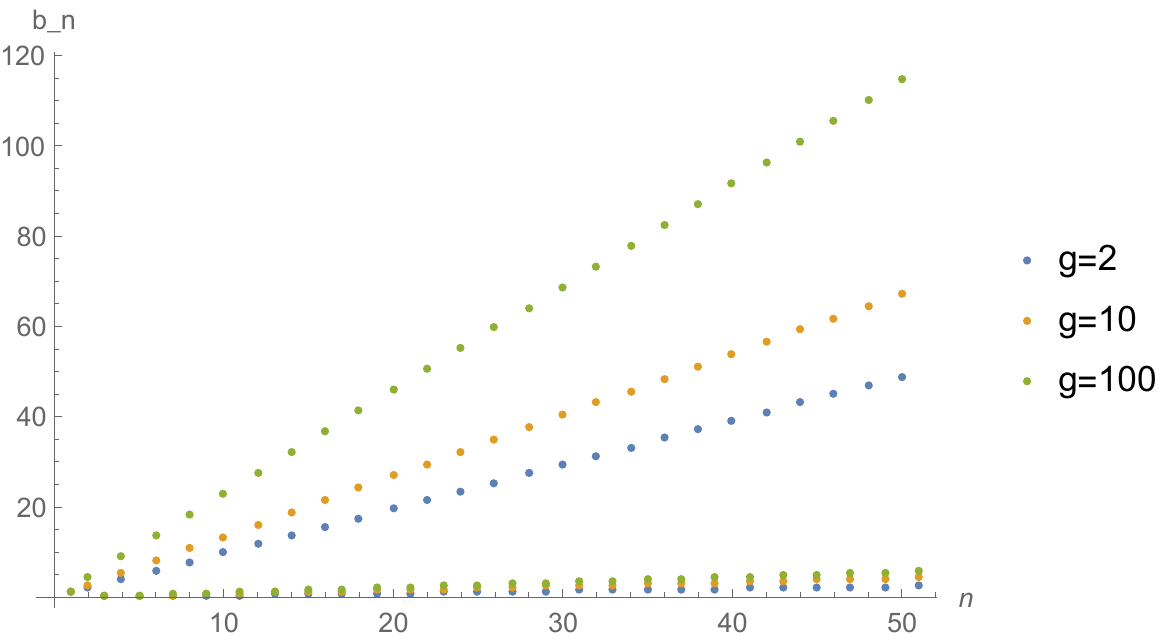}
    \caption{$b_n$}
  \end{subfigure}
  
   \begin{subfigure}[b]{0.70\linewidth}
    \includegraphics[width=\linewidth]{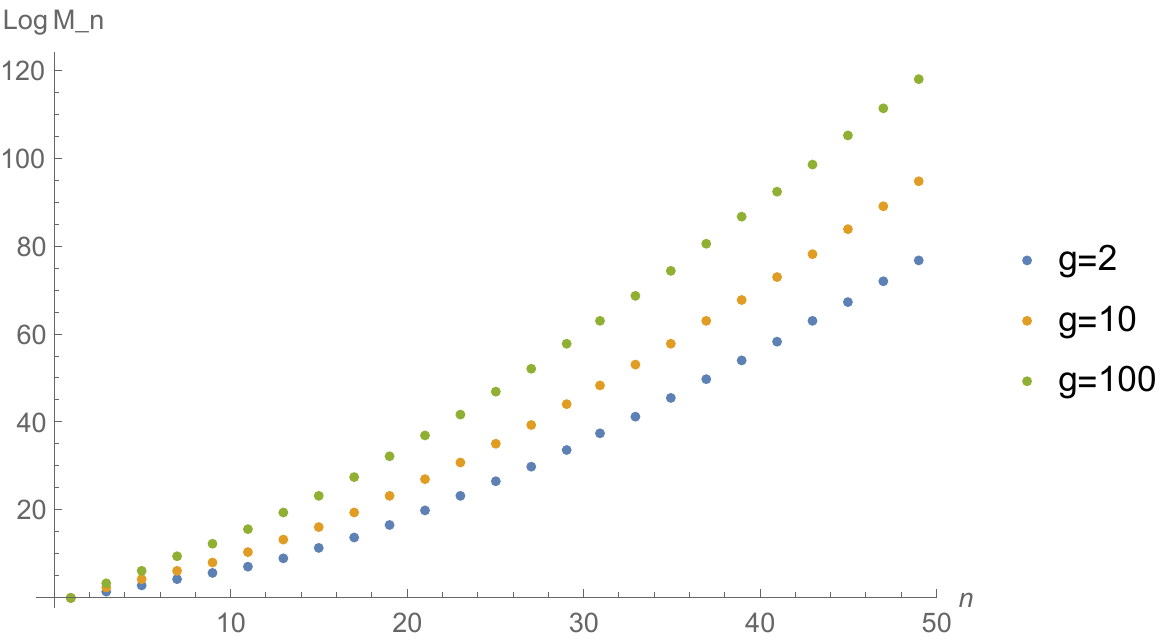}
    \caption{$\log M_n$}
  \end{subfigure}

  \caption{The moments and Lanczos coefficients for different values of $g$.}
  \label{fig9}
\end{figure}

The correlator for the inner product \eqref{inner} at inverse temperature $\beta$ is
\begin{multline}
       C(t, \beta) = \frac{1}{N'} \sum_j \frac{j \pi}{8 L} \frac{e^{\beta \pi j/4L}}{-1 + e^{\beta \pi j /2L} } \\\{ 2 A_j + B_j\}^2 [e^{-i \pi j t/2L} + e^{i \pi j t/2L} ].
\end{multline}
Thus, the moments are given by
\begin{multline}
       M_n = \frac{1}{N'} \sum_j \frac{j \pi}{16 L} \frac{e^{\beta \pi j/4L}}{-1 + e^{\beta \pi j /2L} } \{ 2 A_j + B_j\}^2 \\\Big( \frac{(i \pi j /2L)^n}{i^n}+ \frac{(-i \pi j /2L)^n}{i^n}\Big). 
\end{multline}
As it is obvious from the formula
odd moments are zero and thus the set of the Lanczos coefficients 
\begin{equation}
    a_n =0.
\end{equation}
In Fig. \ref{fig9}, one can see the plots for the set of $b_n$ and the even moments in this case. 
The $ b_n$ coefficients have two linear branches with two different positive slopes. One of the slopes is almost the same for different values of $g$ while another slope increases while the $g$ parameter grows. 
In other words, one slope is a function of $g$ while another one is constant and $g$-independent.
(Look at the example in chapter 4 in \cite{viswanath1994recursion})

\begin{figure}

    \includegraphics[width=\linewidth]{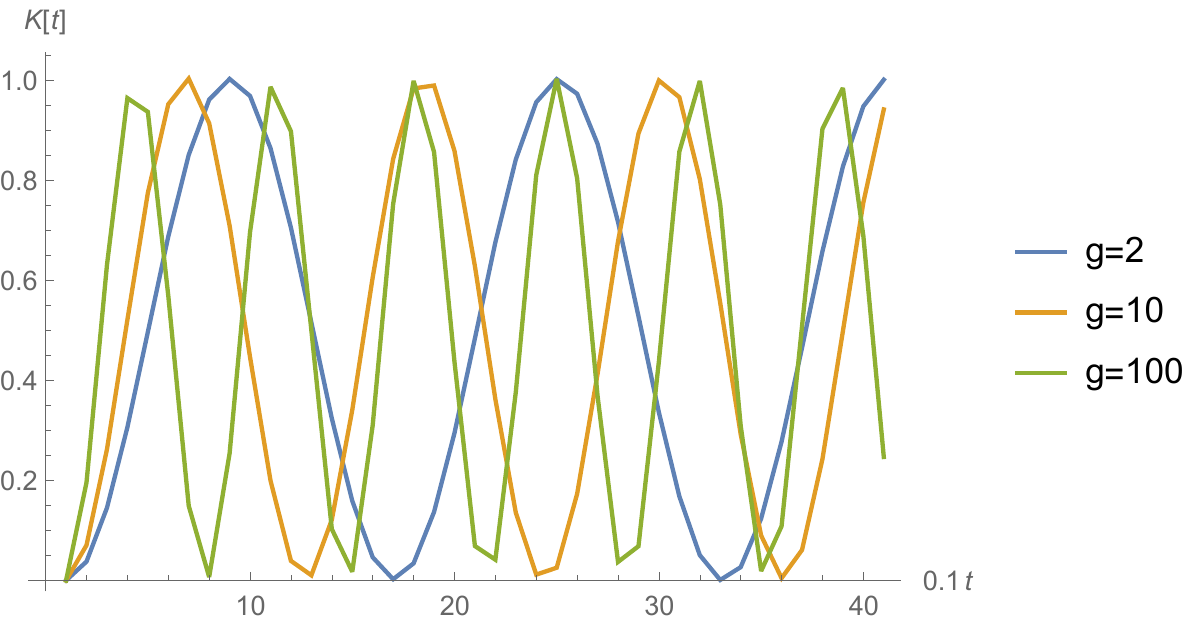}
    \caption{Krylov complexity of the matrix quantum mechanics in the thermal state as a function of time for different values of $g$ at $ \beta = 1$.}

  \label{fig10}
\end{figure}

\begin{figure}
  \centering
  \begin{subfigure}[b]{0.47\linewidth}
    \includegraphics[width=\linewidth]{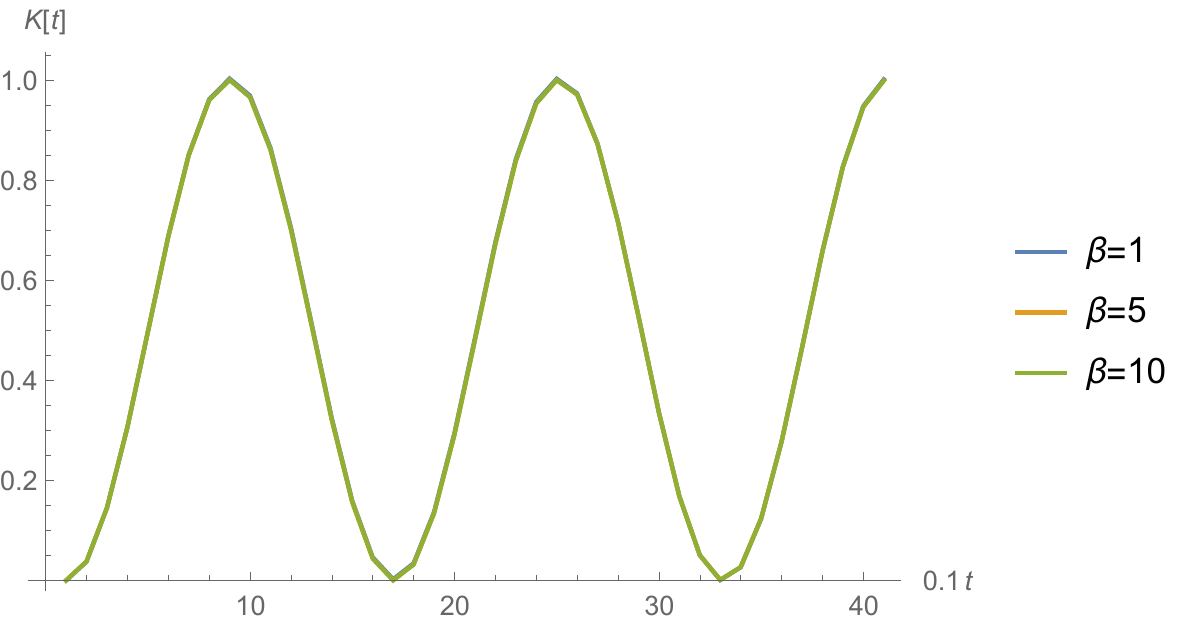}
    \caption{$g = 2$}
  \end{subfigure}
   \begin{subfigure}[b]{0.47\linewidth}
    \includegraphics[width=\linewidth]{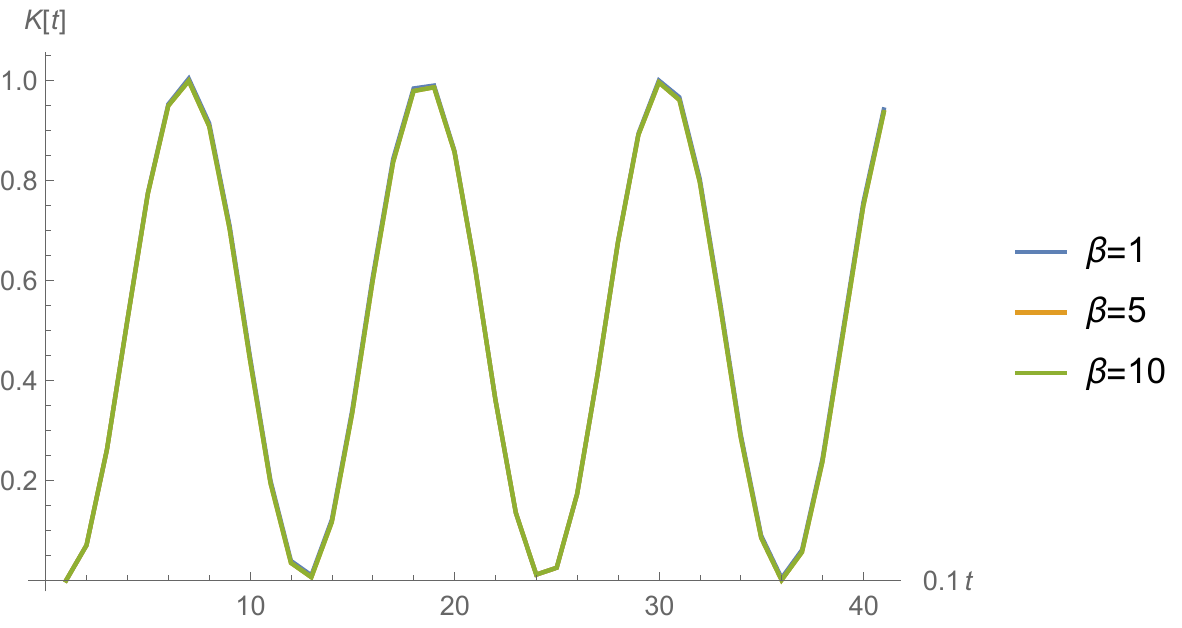}
    \caption{$g = 10$}
  \end{subfigure}

  \caption{Krylov complexity for different values of $\beta$.}
  \label{fig11}
\end{figure}

In Fig. \ref{fig10} and \ref{fig11}, one can see the Krylov complexity of the 1-MQM over the thermal states for different values of $\beta$ and $g$. 
For a fixed $ \beta$, the periodicity of the Krylov complexity decreases as $g$ increases. In this case, unlike over the ground state, the peak of the complexity remains the same for different values of $g$ and $\beta$. Moreover, for the fixed $g$, the periodicity remains the same for different values of $\beta$.

\section{Toda chain flow in Krylov space and radius of convergence of Krylov complexity}

\subsection{Toda chain flow in Krylov space}

This section is mostly based on \cite{dymarsky2019toda}.
They begin by reviewing the basics of the recursion method. First, we start with the time-correlation function of some operator $A$, 
\begin{equation}
    C(t)= \langle A(t), A \rangle 
\end{equation}
it is defined based on the Hermitian form in the space of operators
\begin{equation}
    \langle A, B \rangle \equiv \tr (A^\dagger \rho_1 B \rho_2) = \langle B, A^\dagger \rangle ^*
\end{equation}
here $\rho_1, \rho_2 $ are some hermitian positive semi-definite operators which commute with the Hamiltonian H. 

It is convenient to introduce 
\begin{equation}
    q_n = \ln \langle A_n,A_m \rangle 
\end{equation}
such that 
\begin{equation}\label{11111}
    G_nm = \langle A_n,A_m \rangle = \delta _{nm} e ^{q_n}
\end{equation}
In \cite{dymarsky2019toda}, authors focus on the Euclidean time evolution. 
For a given $O(t)$ where $t$ is Euclidean time, an operator evolved in Minkowski time is $O(-it)$.

The adjoin action of $H$ in the Krylov basis $A_n$ can be represented by Jacobi matrix L, 
\begin{equation}
        [H, A_n] = \sum _m L_{nm} A_m, \qquad L= g M g^{-1}
\end{equation}
\begin{equation}
        g  = diag (e^{q_0/2} , e^{q_1 /2}, ...)
\end{equation}
\begin{equation}
\begin{split}
        M  =  \begin{pmatrix}
        a_0 & b_1 & 0&0& \dots
        \\
        b_1& a_2& b_2 & 0& \dots
        \\
        0& b_2& a_2 & b_3 &\dots
        \\
        0&0& b_3 & a_3 &\dots
        \\
        \vdots & \vdots & \vdots & \vdots & \ddots
    \end{pmatrix}.
\end{split}
\end{equation}
As a generalization of \eqref{11111} we define 
\begin{equation}
    G_{nm} (t) = \langle A_n (t),A_m \rangle
\end{equation}
and evolution in terms of the Lanczos coefficient 
\begin{equation}
    G(t) = g e^{Mt} g^T
\end{equation}
the original correlation function is then 
\begin{equation}
    C(t) = G_{00} (t) = \langle A_0, A_0 \rangle (e^{Mt})_{00}
\end{equation}
Lanczos coefficients $a_n, b_n$ can be promoted to be t-dependent. 

Therefore, we can apply the recursion method to define the Krylov basis starting from the same initial $A$ for any given value of t. 
This defines the orthogonal basis $A_n ^t, A_0 ^t \equiv A$, 
\begin{equation}
    G_nm^t \equiv \langle A_n^t, A_m^t \rangle _t = \delta _{nm} e^{q_n(t)}
\end{equation}
where $a_n(t), b_n(t)$ and $q_n$ are now t-dependent and thus $M(t)$ and $g(t)$ are time-dependent as well.

An important observation is that $G_{nm}(t)$ and $G^t _{nm}$ written in terms of two different bases $A_n, A_n^t$. 
They are related by a change of coordinates
\begin{equation}
    \begin{split}
        G(t) &= z(t) G^t Z(t)^T
        \\
        A_n & = \sum _m z_nm (t) A_m^t.
    \end{split}
\end{equation}
The basis $A_n^t$ has been transformed into basis $ A_n = A_n^{t=0}$ by the matrix $z(t)$.
One can express $G^t$ in terms of $ g(t)$
\begin{equation}
    G(t) = g(0) e^{M(0) t} g(0)^T = Z(t) g(t) g(t)^Tz(t)^T.
\end{equation}
Explicit time dependence of G(t) provides that 
\begin{equation}
    \frac{d}{dt}(G^{-1} \dot{G}) =0 .
\end{equation}
It follows that $ q_n(t)$ satisfies the Toda equation. The relation between $a_n, b_n$, and $q_n$ is given by 
\begin{equation}
a_n(t) \equiv \dot{q}_n
\end{equation}
\begin{equation}
     b_n(t) \equiv e^{(q_{n+1} - q_n)/2}.
\end{equation}
One can introduce $\tau _n = \exp (\sum _{0 \leq k \leq n } q_n)$. 
In particular
\begin{equation}
    \tau_0(t) = e^{q_0(t)} = C(t)
\end{equation}
which says that the time-correlation function analytically continued to Euclidean time is a tau-function of the Toda hierarchy. 

Furthermore, since $ e^{- q_n(0) /2 A_n}$ and $  e^{- q_n(t) /2 A_n^t(t/2)}$ are orthonormal bases, they must be related by an orthogonal transformation $Q^{T}$ ( for more detail look at \cite{dymarsky2019toda}) 
\begin{equation}
    \sum_m Q^T_{nm}(t/2) e^{q_m(0)/2} A_m = e^{-q_m(t)/2} A_n^{t}(t/2) 
\end{equation}
Evolving this equation in time by $-t/2$. We find 
\begin{equation}
    e^{M(0) t } = Q(t) R(t), \qquad R^T (t/2) = g^{-1}(0) Z(t) g(t).
\end{equation}
This QR decomposition of $ e^{M(0) t}$ \cite{symes1980hamiltonian}.

In \cite{dymarsky2019toda}, the authors apply the relation of Lanczos coefficients to the Toda chain to clarify chaos in quantum many-body systems. 
An accurate counting of nested commutators appearing in the Taylor series expansion of $ C(t)$ will be singular at some finite $ t = t^*$.

In general, chaotic behavior is reflected by the linear growth of both $a_n$ and $b_n$. While the slope of $a_n$ can not exceed twice the slope of $b_n$. 

To study the singular behavior of the time-correlation function, we assume that $ C(t) = G_{00} (t)$ together with its derivatives are smooth functions for $ 0 \leq t< t^*$, and diverges at $ t=t^*$. 
From here follows that $G_{nm}(t)$ are regular for $0\leq t < t^*$. 
Using QR decomposition, we have
\begin{equation}
    R_{00}(t/2)^2 = C(t)/ C(0)
\end{equation}
and conclude that $R_{00}(t)$ is regular for $ 0\leq t < t^* /2 $ and diverge at $t= t^*/2$. 
We can decompose $A(t)$ into orthogonal Krylov basis
\begin{equation}
    e^{q_0/2} A(t) = \sum _n \phi_n(t) (  e^{q_0/2} A_n )
\end{equation}
while 
\begin{equation}
    \phi_n(t) = R_{00} (t)  Q_{n0} (t)
\end{equation}
This is a manifestation of delocalization in Krylov space. At $t=t^*/2$, the operator $A(t)$ spreads across the whole Krylov space. 

Just note that this singularity is along the imaginary axis as we consider the Euclidean time.

\subsection{Radius of convergence of the Krylov complexity}

The Krylov complexity is defined in \eqref{22222} can be written in terms of $\phi_n$. 
Thus, in the case that in our calculation, the coefficients $\phi_n$ are regular in the time-band 
$0\leq t \leq t^*/2$, the Krylov complexity is also regular in this time band and the radius of convergence for Krylov complexity is $t^*/2$, 
while $t^*$ is the radius of convergence of the correlation function.

Now in the 1-MQM model we consider in this project, first we should find the radius f convergence of the correlator.  

For a given series $ \sum_n f_n$, the series converge if
\begin{equation}
    \lim _{n \rightarrow \infty } \Big|\frac{f_{n+1}}{f_n}\Big| < 1.
\end{equation}
In the case of 1-MQM over the ground state we have 
\begin{equation}
    f_n = \frac{-i\pi}{8 L^2} n ( 2 A_n + B_n)^2 e^{-i \pi t n / 2L}
\end{equation}
therefore 
\begin{equation}
    \lim _{n \rightarrow \infty } \Big|\frac{f_{n+1}}{f_n}\Big| = \frac{n+1}{n} e^{-i \pi t/ 2L} \Big( \frac{2A_{n+1}+ B_{n+1}}{2A_n + B_n}\Big)^2 < 1.
\end{equation}
After analytically continuation of t, for a complex $ z= x+iy$, one can get 
\begin{equation}
   |e^{-i\pi z /2L}|=  |e^{\pi y / 2L}| 
\end{equation}
Therefore
\begin{equation}
    y < \frac{2L}{\pi} \lim _{n \rightarrow \infty}  \ln \Big( \frac{n}{n+1} \big( \frac{2 A_n + B-n}{2 A_{n+1} + B_{n+1}}\big)^2\Big) 
\end{equation}
Thus the radius of convergence of correlation function $ C(t)$ is at most
\begin{equation}
     t^* = \frac{2 L}{\pi} \lim _{n \rightarrow \infty }  \ln \Big( \frac{n}{n+1} \big( \frac{2 A_n + B-n}{2 A_{n+1} + B_{n+1}}\big)^2\Big) 
\end{equation}

In Fig. \ref{fig8}, one can see that the first peak of the Krylov complexity for all values of $g$ is approximately at $ t^* /2$. The radius of convergence of the Krylov complexity is discussed in the previous part.   

\section{Discussion}

In this paper, we investigate the notion of Krylov complexity for 1-MQM which is a toy model of the gauge theory dual of an AdS black hole in both ground states and thermal states.  In both cases, we observe the linear growth. However, over the thermal states, it divides into odd and even branches.  
This work is a warm-up for the more realistic examples of holography as BFSS. Even studying the n-matrix quantum mechanics can be much more challenging as the system is not solvable, thus, we treat the chaotic systems. Since the theory is not solvable, it is a very hard task to find the correlation functions in the theory.  
Here albeit the 1-MQM is not chaotic, the Lancsoz coefficients have the linear growth behavior. However, in the literature, it has been conjectured that the linear growth of the Lancsoz coefficients is a sign of chaos \cite{parker2019universal}. It seems that the conjecture is not universal and it should be corrected in a way that only if we have a chaotic system, the corresponding Lancsoz coefficients have linear growth. 
Furthermore, in the case of the thermal state both in the harmonic oscillator and 1-MQM we see that the $b_n$ coefficients are divided in the odd and even branches with linear growth. The reason why this happens is not clear but it might be because of using the new inner product \eqref{inner} as it is the case also in \cite{avdoshkin2022krylov}. Moreover, it remains an open question: if it is possible in some scenarios that we have even more than 2 branches. 
The result that we find for the complexity is periodic since the correlators are periodic. However, we calculate the radius of convergence of the Krylov complexity where it is half of the radius of convergence of the correlator. From the numerics, it is obvious that  
this time is the same point as the first peak of the Krylov complexity.  Thus, in the radius of convergence we see only the growth of Krylov complexity.  Nevertheless, finding the Krylov complexity after this point is an open question.

\emph{Acknowledgements.--}
I would like to thank M. Alishahiha for useful discussions and communication. In particular, I would like to thank my supervisor, K. Papadodimas for the useful discussion and comment on the draft. I would like also to thank my co-supervisor  M. Brtolini for his support during this work. I would like to thank CERN-TH for its hospitality during the preparation of this work.
The research is partially supported by the INFN Iniziativa Specifica- String Theory and Fundamental Interactions project.

\appendix

 \section{Collective field theory formalism}

Broadly, the collective approach involves a variable transformation. Consider an operator Hamiltonian
\begin{equation}
    \hat{H} = \frac{1}{2} \sum _{i=1}^{N} P_i^2 + V (q_1, ..., q_M) 
\end{equation}
in a manner that allows its representation using an infinite set of new variables
\begin{equation}
  \phi (x) = f(x, q_1, ..., q_M).  
\end{equation}
This set would be generally over-complete for finite $M$. One can make a standard canonical transformation and express the theory using $ \phi(x)$.
Thus, the wave function of the theory should be written in terms of $\phi(x) $.
this can come about as a restriction on invariant singlet subspace of the full Hilbert space. On the wave functional, the kinetic term takes the form
\begin{multline}
        K\equiv - \frac{1}{2} \sum \frac{\partial ^2}{\partial q _i^2} = \frac{1}{2} \int dx~ \omega (x, \phi) \frac{\delta }{\delta \phi(x)} - \frac{1}{2} \\\int dx dy ~ \Omega (x, y, \phi) \frac{\delta }{\delta \phi(x)} \frac{\delta }{\delta \phi(y)} 
\end{multline}
where 
\begin{equation}
    \begin{split}
         \omega (x, \phi) & = \sum _i \partial_i^2 f(x,q)
         \\
         \Omega (x, y, \phi)& =\sum_i \partial_i f(x,q) \partial_if(y,q) 
    \end{split}
\end{equation}
The kinetic term in the new collective representation is not Hermitian. It is because of the fact that the new scalar product involves a Jacobian.

Using a similar transformation, one finds the following Hermitian Hamiltonian
\begin{equation}
H = \frac{1}{2} \Pi~ \Omega ~\Pi + \frac{1}{8} (\omega + \frac{\partial \Omega }{\partial \phi }) \Omega^{-1} (\omega + \frac{\partial \Omega }{\partial \phi }) + V[\phi] - \frac{1}{4} \frac{\delta \omega}{ \delta \phi} - \frac{1}{4} \frac{\partial^2 \Omega}{ \partial \phi \partial \phi} 
\end{equation}
whith $ \phi(x)$ and $ \Pi (x)$ being a conjugate set of fields variables when $ \Pi (x) = - i \delta/ \delta \phi (x)$

\bibliographystyle{apsrev4-2} 
\bibliography{references.bib} 

\end{document}